\documentclass{article}
\usepackage{arxiv}
\usepackage[utf8]{inputenc} 
\usepackage[T1]{fontenc}    
\usepackage[hidelinks]{hyperref}       
\usepackage{url}            
\usepackage{booktabs}       
\usepackage{amsfonts}       
\usepackage{nicefrac}       
\usepackage{microtype}      
\usepackage{amsmath}
\usepackage{listings}
\usepackage{listingsutf8}
\usepackage{graphicx}
\usepackage{algorithm}
\usepackage{algpseudocode}
\usepackage{xcolor}
\definecolor{LightGray}{gray}{0.95}
\usepackage[frozencache=true,cachedir=.]{minted}
\setminted[python]{
    linenos=false,
    breaklines=true,
    encoding=utf8,
    fontsize=\footnotesize,
    bgcolor=LightGray,
}

\usepackage{etoolbox,xpatch}

\makeatletter
\AtBeginEnvironment{minted}{\dontdofcolorbox}
\def\dontdofcolorbox{\renewcommand\fcolorbox[4][]{##4}}
\xpatchcmd{\inputminted}{\minted@fvset}{\minted@fvset\dontdofcolorbox}{}{}
\xpatchcmd{\mintinline}{\minted@fvset}{\minted@fvset\dontdofcolorbox}{}{} 
\makeatother

\usepackage{ulem}
\usepackage{natbib}
\DeclareUnicodeCharacter{03B1}{$\alpha$}
\DeclareUnicodeCharacter{03B2}{$\beta$}
\DeclareUnicodeCharacter{03BC}{$\mu$}
\DeclareUnicodeCharacter{03C3}{$\sigma$}
\DeclareUnicodeCharacter{03B8}{$\theta$}

\title{Bayesian Additive Regression Trees for Probabilistic programming}

\author{
  Miriana Quiroga \\
  IMASL-CONICET\\
   \And
  Pablo G Garay \\
  IMASL-CONICET\\
   \And
  Juan Martin Loyola \\
  IMASL-CONICET\\
  \AND
  Juan M. Alonso \\
  IMASL-CONICET\\
  \And
  Osvaldo A Martin \\
  IMASL-CONICET-UNSL\thanks{Av. Italia 1556, San Luis. Argentina.}\\
  \texttt{omarti@unsl.edu.ar}\\
}

\begin{document}
\maketitle

\begin{abstract}

Bayesian additive regression trees (BART) is a non-parametric method to approximate functions. It is a black-box method based on the sum of many trees where priors are used to regularize inference, mainly by restricting trees' learning capacity so that no individual tree is able to explain the data, but rather the sum of trees. We discuss BART in the context of probabilistic programming languages (PPL), i.e., we present BART as a primitive that can be used as a component of a probabilistic model rather than as a standalone model. Specifically, we introduce the Python library PyMC-BART, which works by extending PyMC, a library for probabilistic programming. We showcase a few examples of models that can be built using PyMC-BART, discuss recommendations for the selection of hyperparameters, and finally, we close with limitations of our implementation and future directions for improvement.
\end{abstract}


\section[Introduction]{Introduction} \label{sec:intro}

Bayesian Additive Regression Trees introduced by~\citet{Chipman2010}, have been demonstrated to be competitive compared with alternatives such as Gaussian processes, random forests, or neural networks. The main reason is that BART needs minimal user input and tuning while maintaining a good performance~\citep{Chipman2010, Rockova2018, bart_review, bart_r_package}. Additionally, and similar to other probabilistic methods like Gaussian processes, BART provides uncertainty quantification via probable intervals.

BART has been applied to solve numerous applications in recent years including estimation of causal effects~\citep{leonti2010causal,hill2011bayesian, hu_causal_2022, Steele2022, chen2022}, species distribution modelling~\citep{embarcadero}, estimating indoor radon concentrations~\citep{kropat2015improved}, modeling of asteroid diameters~\citep{de_Souza_2021}, genomics~\citep{li_genomic_2022}, just to name a few.

Variable selection has also been an area of interest by BART researchers, achieving reasonable results~\citep{Bleich2014, linero2018bayesian}. The main approach is based on counting how many times a given covariable is incorporated into the trees relative to the other covariables in the same model. In order to improve the performance of this very simple variable selection procedure,~\citet{linero2018bayesian} introduced a sparsity-inducing Dirichlet hyperprior on the splitting proportions of the regression tree prior.

In the literature, it is common to find specific BART models associated with specific samplers and thus specific implementations \citep{Chipman2010, pratola_2020, Lamprinakou_2020, Orlandi_2021}. In other words, general methods to sample from BART posteriors are not common and instead, conjugate priors are used to define such BART models. Likewise, packages like BART R \citep{bart_r_package}, Xbart \citep{xbart} or bartpy \citep{bartpy}  have been designed as standalone packages restricted to predefined models and not as tools to build arbitrary probabilistic models with BART components.

In this work, we present PyMC-BART an implementation of BART that removes the conflation of inference and modeling. As far as we know, this is the first package introducing BART as a primitive for a probabilistic programming language, and we are only aware of two previous proposals related to generalizing BART. Namely, \citet{general_bart} presented the general BART framework unifying BART extensions that were previously presented as separated models and~\citet{linero2022} introduced a reversible jump Markov chain Monte Carlo algorithm to bypass the need for conjugacy. Abstracting away the implementation of inference from modeling, as we do with PyMC-BART, has been key to the success of applied Bayesian modeling in recent years, and the main reason for the popularity of PPLs like PyMC~\citep{pymc2023}. 

As previously mentioned, PyMC-BART works by extending PyMC's functionality. From the user's perspective, the main new object introduced is a BART random variable that can use together with the built-in PyMC random variables to build arbitrary probabilistic models in PyMC. We also provided a new inference method\footnote{or step methods, using the PyMC nomeclature} called PGBART, which works exclusively for BART variables. For most users, there is no need to directly interact with PGBART. Instead, it is designed in such a way that PyMC will automatically use it whenever a BART variable occurs as part of a model. Thus, other than using a BART variable in a model, all other modeling aspects remain the same from a user perspective. Including model building, conditioning on observed data, fitting the model, computing posterior predictive samples, etc. Even more, due to the tight integration of PyMC with ArviZ, tasks such as assessing the fit, comparing to other models, assessing convergence, and other exploratory analysis of Bayesian models' tasks~\citep{Kumar2019, Martin2021} are immediately accessible to users of PyMC-BART without the need to learn a new set of tools or syntax. Finally, we provide a set of tools specifically designed to work with the posterior distribution of trees, including tools to aid in the interpretation of the fitted function, to help users perform variable selection, and also to help diagnose the MCMC samples.

For the rest of this article, we will focus primarily on the practical aspects of BART. We start by giving a very brief overview of the BART model in Section~\ref{sec:overview}, then we continue with Section~\ref{sec:api} describing the  PyMC-BART's API, including which hyperparameters are available to  users. In Section~\ref{sec:examples} we demonstrate basic usage through examples. And in Section~\ref{sec:loo} we discuss how to choose the number of trees for BART, arguably the most important hyperparameter PyMC-BART users can change. Finally, we conclude with Section~\ref{sec:discussion} discussing some limitations of the current state of our implementation and the future of PyMC-BART. While the target audience of this manuscript is practitioners interested in adding BART to their Bayesian toolkit, we also provide details of the sampler we use to approximate the posterior over the trees in Appendix~\ref{sec:pgbart}, which we hope will help others interested in contributing to the base code.

PyMC-BART is available from the Python Package Index at \url{https://pypi.org/project/pymc-bart} and it can also be installed using conda. The package documentation, including installation instructions and examples of how to use BART to conduct different statistical analyses, can be found at \url{https://www.pymc.io/projects/bart}.

The version of PyMC-BART used for this article is 0.5.1 and the version of PyMC is 5.1.2 \citep{pymc2023}. All analyses are supported by extensive documentation in the form of interactive Jupyter notebooks~\citep{Kluyver2016} available in the paper repository on GitHub \url{https://github.com/Grupo-de-modelado-probabilista/BART/tree/master/experiments}, enabling readers to re-run, modify, and otherwise experiment with the models described here on their own machines. This repository also includes instructions on how to set up an environment with all the dependencies used when writing this manuscript.

\section{The BART model} \label{sec:overview}

\begin{figure}[t!]
    \centering
    \includegraphics[width=0.8\textwidth]{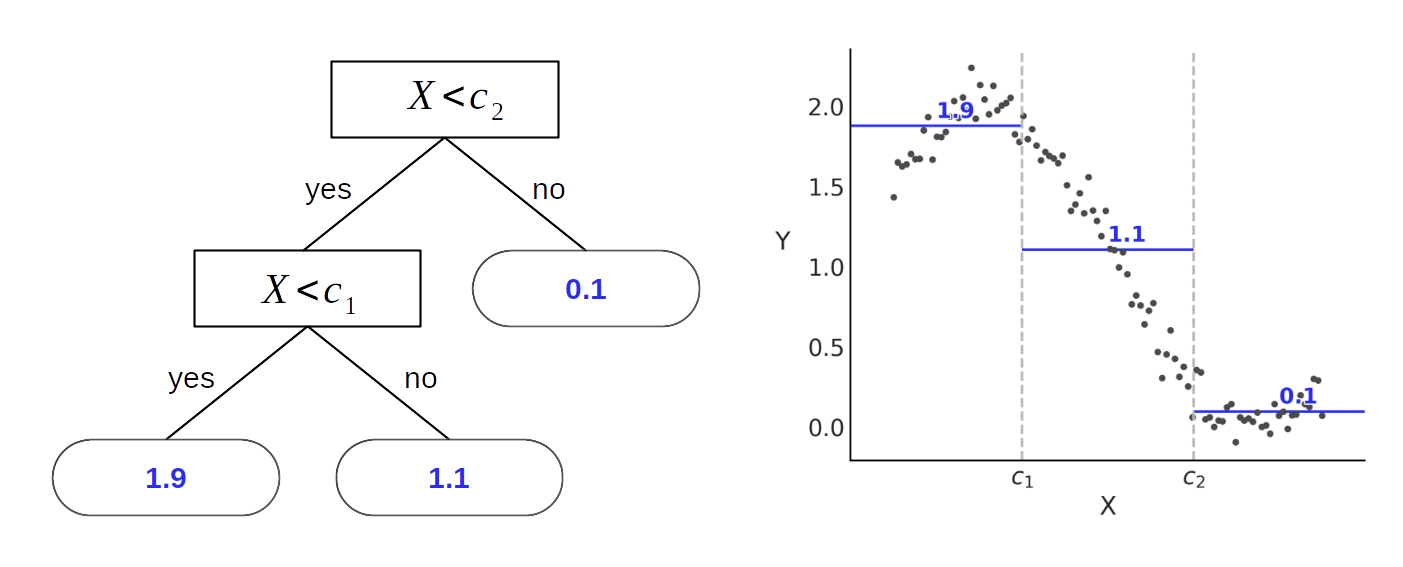}
     \caption{Example of a regression tree (left) and the induced partition of the data space (right). The interior nodes are represented with rectangles. Inside them, we can find the splitting variable $X$ and the splitting values $c_1$ and $c_2$. The terminal nodes are represented using rounded rectangles, inside them, we find the leaf values in blue. In this example, we show a single tree fitting the data. For BART we use a sum of trees.}
     \label{fig:reg_tree}
\end{figure}

A BART model can be represented as:
\begin{equation}
Y \sim \phi \left(\sum_{i=1}^m G_i(\boldsymbol{X}; \mathcal{T}_i, \mathcal{M}_i), \theta \right)
\label{eq:bart}
\end{equation}
\noindent where $\boldsymbol{X}$ are the covariates and $Y$ the response variable. The sum is done over $m$ trees. Each $G_i$ is a binary tree with structure, $\mathcal{T}_i$ i.e., $\mathcal{T}_i$ specifies the set of interior nodes (also known as splitting nodes) and the associated splitting rules, and the set of terminal nodes (also known as leaf nodes). $\mathcal{M}_i = \{\mu_{1,i}, \mu_{2,i}, \cdots, \mu_{b, i} \}$ represents the values at the terminal nodes. For an example of a single tree used for regression, see Figure~\ref{fig:reg_tree}. $\phi$ represents an arbitrary probability distribution and $\theta$ parameters of $\phi$ that are not modeled as a sum of trees, like the standard deviation for a Normal likelihood.

The model is completed by specifying priors for $\mathcal{T}$ and $\mathcal{M}$. For $\mathcal{T}$, independent priors are set for the depth of the trees, the splitting variables, and the splitting values. Details for such priors can be found in Appendix~\ref{sec:pgbart}. The overall effect of the BART priors is to prevent overfitting by making trees shallow, making leaf node values small on the scale of the data, and regularizing statistical interactions.

\begin{figure}[t!]
    \centering
    \includegraphics[width=0.8\textwidth]{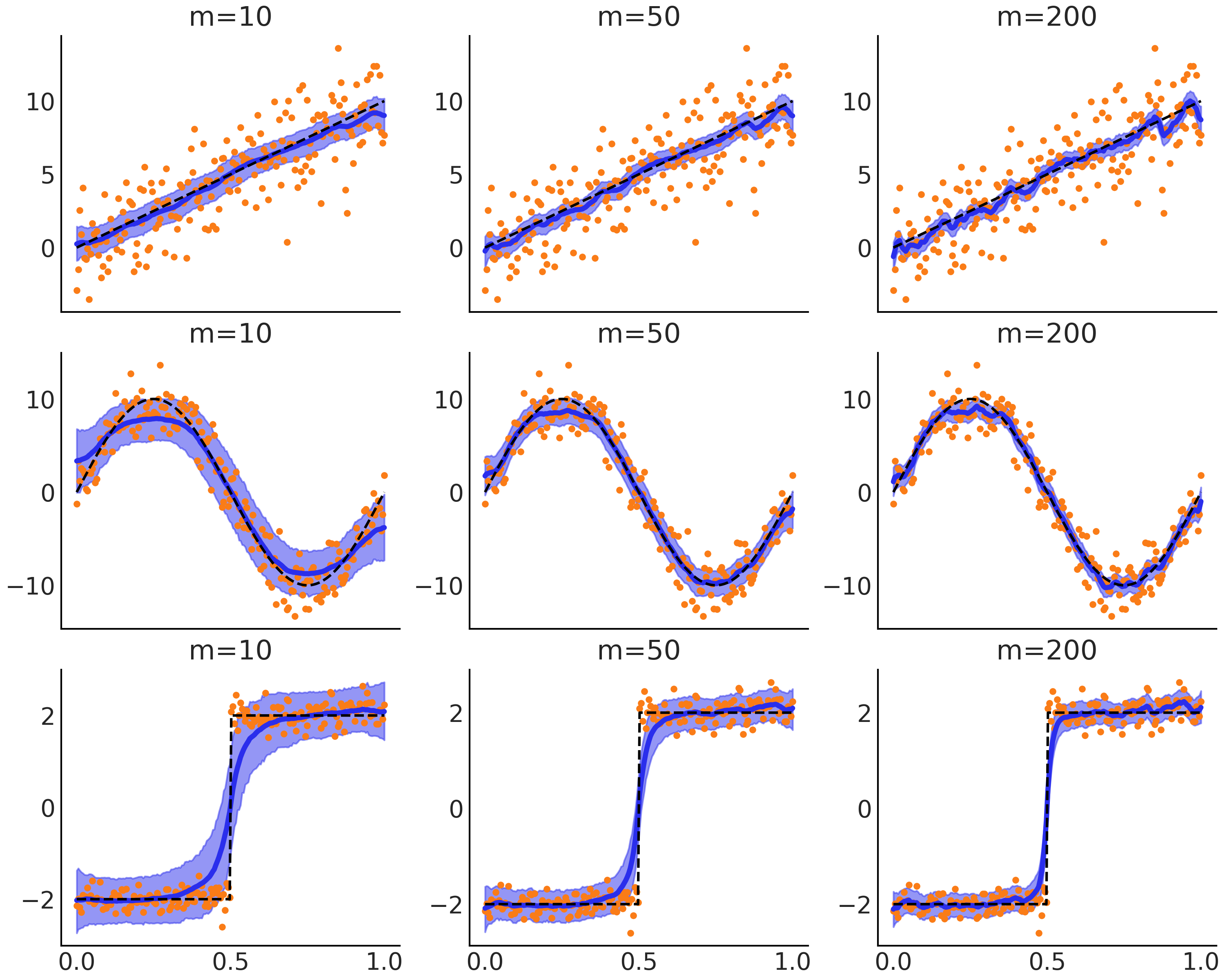}
     \caption{Examples of a BART model fitted to data (orange dots) from 3 different functions (black dashed lines) with  \texttt{m}=10 (left column),  \texttt{m}=50 (center column) or  \texttt{m}=200 (right column). The mean predicted function and its HDI are shown in blue and light blue, respectively.}
     \label{fig:simple_functions}
\end{figure}

We can think of BART as priors over step functions, i.e., priors over piecewise constant functions. Furthermore, in the limit of the number of trees $m \to \infty$, BART converges to a nowhere-differentiable Gaussian Process \citep{Linero2018smoothness}. One may object that BART cannot directly model smooth functions, which are arguably the most common cases for most datasets. Still, BART is useful in practice, as judged by all its applications. Figure~\ref{fig:simple_functions} shows an example of BART fitted to data generated from 3 simple functions, a line, a sine, and a step function. In all the examples, the sample size is 200. We can see the effect of \texttt{m} on the result, the HDI (blue band) becomes narrower, and the mean predicted function better fit the curves, in Section~\ref{sec:loo} we provide some guidance on how to select \texttt{m}. 

BART is part of the ensemble models family. Members of this family are based on modeling a function and making predictions, from a combination of simple models instead of from a single complex one. In the case of BART, the simple models are the binary trees. Generally, ensemble models have desirable properties, like being less prone to over-fit~\citep{zhou_ensemble_2012, kuhn_applied_2013}. A drawback of ensemble methods, as well as other black-box methods, is that a meaningful and direct interpretation of the fitted model is difficult or even impossible; instead, models are assessed via evaluations at given values of the covariates. Later, we present examples of partial dependence plots~\citep{friedman_2001}, which can be efficiently computed with PyMC-BART and can aid users in the interpretation of the results from BART models.

\section{API}
\label{sec:api}

PyMC-BART seamlessly integrates with PyMC and ArviZ; for example, priors for $\theta$ in Equation~\ref{eq:bart}, i.e., non-BART related parameters, can be arbitrarily set by the user using the standard PyMC syntax. The same goes for fitting the model: users need only to call the  \texttt{pm.sample()} function, as they would normally do for regular models.

Additionally, PyMC uses ArviZ's InferenceData object~\citep{Kumar2019, arviz} to store posterior samples, prior/posterior predictive samples, sampler statistics, etc. (see Figure~\ref{fig:inf_data}). InferenceData is a rich data structure based on Xarray~\citep{xarray_2017}. For PyMC models with a BART variable from PyMC-BART, the InferenceData object will also store the variable inclusion record. This data can then be processed by helper functions provided by PyMC-BART, to obtain variable importance plots, as we show in Section~\ref{sec:examples}. Users can also use ArviZ for convergence diagnostics including $\hat R$, effective sample size, trace plots, and rank plots \citep{10.1214/20-BA1221, Martin2021}.

In the text, we assume the PyMC package has been imported with the  \texttt{pmb} alias.

We now discuss the 3 main elements provided by PyMC-BART :

\begin{itemize}
    \item A BART random variable. This behaves similarly to default PyMC random variables and can be used together with them, allowing users to build arbitrary probabilistic models. 
    \item The PGBART step method. A method to sample from a BART random variable. The user does not need to interact with this sampler, as PyMC will automatically assign it to a BART variable if such a variable is part of the model.
    \item A collection of helper functions to aid the user interpret the results.
\end{itemize}

\subsection{BART random variables}
For those familiar with PyMC syntax, defining BART models is straightforward. A  \texttt{pmb.BART} random variable behaves like other PyMC variables, with a few caveats. It requires two mandatory arguments:  \texttt{X}, a 2D NumPy array or a pandas DataFrame representing the covariate matrix, and  \texttt{Y}, a 1D NumPy array or a pandas  DataFrame representing the response vector.  \texttt{Y} is needed because PyMC-BART uses it to define the starting point when sampling from the BART posterior and the variance of the leaf nodes.

Another difference is that a  \texttt{pmb.BART} variable does not accept other PyMC random variables as arguments. The main reason is that, as in other implementations, the priors for the BART variables are not directly set by the users, but instead hyperparameters are used to adjust them indirectly\footnote{While this seems to work in practice, extensions could be considered in the future, like using a prior over $m$, instead of a fixed number.}. The hyperparameters that can be changed by the user are:

\begin{itemize}
    \item The number of trees  \texttt{m}. This is a positive integer that defaults to 50. For some datasets, values as low as 20 could provide a good approximation; for others, values as high as 200 may be needed. The value of  \texttt{m} can be defined using cross-validation. In our experiments, we found that Pareto smoothed importance sampling leave-one-out cross-validation (PSIS-LOO-CV)~\citep{vehtari_2017, vehtari_2021}, can be used to find reasonable values of  \texttt{m} (see Section~\ref{sec:loo}).
    \item The value of  \texttt{alpha} and   \texttt{beta}, controlling the node depth. These parameters are the same as those originally proposed by~\citet{Chipman2010}. 
    \item The prior over the splitting variables. This is uniform over the covariates  \texttt{X}. Users can pass an array, of the same length as the total number of covariates, if they have prior information about the relative importance of the variables.
    \item The  \texttt{response}. How the leaf node values are computed. Available options are constant (a single value at each leaf node), linear (a linear regression) or mix (both options selected at random). Defaults to constant. Options linear and mix are still experimental.
    \item \texttt{split\_rules}. Allows using different split rules for different columns. The default is \texttt{ContinuousSplitRule}. The two other options are  \texttt{OneHotSplitRule} and  \texttt{SubsetSplitRule} \citep{deshpande2023}, both meant for categorical variables. 
    \item  \texttt{shape}. Specify the shape of the response vector. Defaults to the length of $Y$. It can be used to model more than one BART random variable per model. By default, the tree structure is shared for all BART random variables but the leaf values are independent. This can be changed by  \texttt{separate\_trees} hyperparameter.
    \item  \texttt{separate\_trees}. Default to False. When training multiple trees (by setting a shape parameter), the default behavior is to learn a joint tree structure and only have different leaf values for each. This flag forces a fully separate tree structure to be trained instead. This is unnecessary in many cases and is considerably slower, multiplying run-time roughly by the number of dimensions.
\end{itemize}

\subsection{PGBART}
PyMC is capable of automatically assigning different sampling algorithms to different parameters in the same model. Thus, PyMC will use the  \texttt{pmb.PGBART} sampler for BART variables, and use one of the built-in PyMC samplers for $\theta$. If $\theta$ is a continuous parameter, then PyMC will automatically choose the  \texttt{pm.NUTS} sampler~\citep{Hoffman2014}.  \texttt{pmb.PGBART} is a sampler we have specifically developed for BART variables, see Appendix~\ref{sec:pgbart} for details. These are the hyperparameters related to the sampler:

\begin{itemize}
    \item  \texttt{num\_particles}. The number of particles used to sample a new tree. Defaults to 10. In cases where the $\hat R$ values~\citep{10.1214/20-BA1221} are too high, increasing the number of particles can help.
    \item  \texttt{batch}. Number of trees out of the  \texttt{m} trees fitted per step. Defaults to  "auto", which is 10\% of  \texttt{m} during and after tuning. Users can provide a tuple, with the first element being the batch size during tuning and the second the batch size after tuning. Increasing  \texttt{batch} can help to reduce $\hat R$ but in our experience, it is better to increase the number of particles.
\end{itemize}

\subsection{Helper functions}

PyMC-BART  currently offers 4 utility functions.

\begin{itemize}
    \item  \texttt{pmb.plot\_pdp}: A function to generate partial dependence plots~\citep{friedman_2001}. Its main inputs are a BART random variable (once a model has been fitted), and the covariate matrix.
    \item  \texttt{pmb.plot\_ice}: A function to generate individual conditional expectation plots~\citep{Goldstein2013PeekingIT}. Its main inputs are a BART random variable (once a model has been fitted), and the covariate matrix.
    \item  \texttt{pmb.plot\_variable\_importance}: A function to estimate the variable importance. Its main inputs are an InferenceData object containing the  \texttt{"variable\_inclusion"} record in  \texttt{sample\_stats} group, a BART random variable (once a model has been fitted), and the covariate matrix.
    \item  \texttt{pmb.plot\_convergence}: A function that plots the empirical cumulative distribution for the effective sample size and $\hat R$ values for the BART random variables. Its main inputs are InferenceData and the name of the BART variable.
\end{itemize}

\begin{figure}
    \centering
    \includegraphics[width=0.8\textwidth]{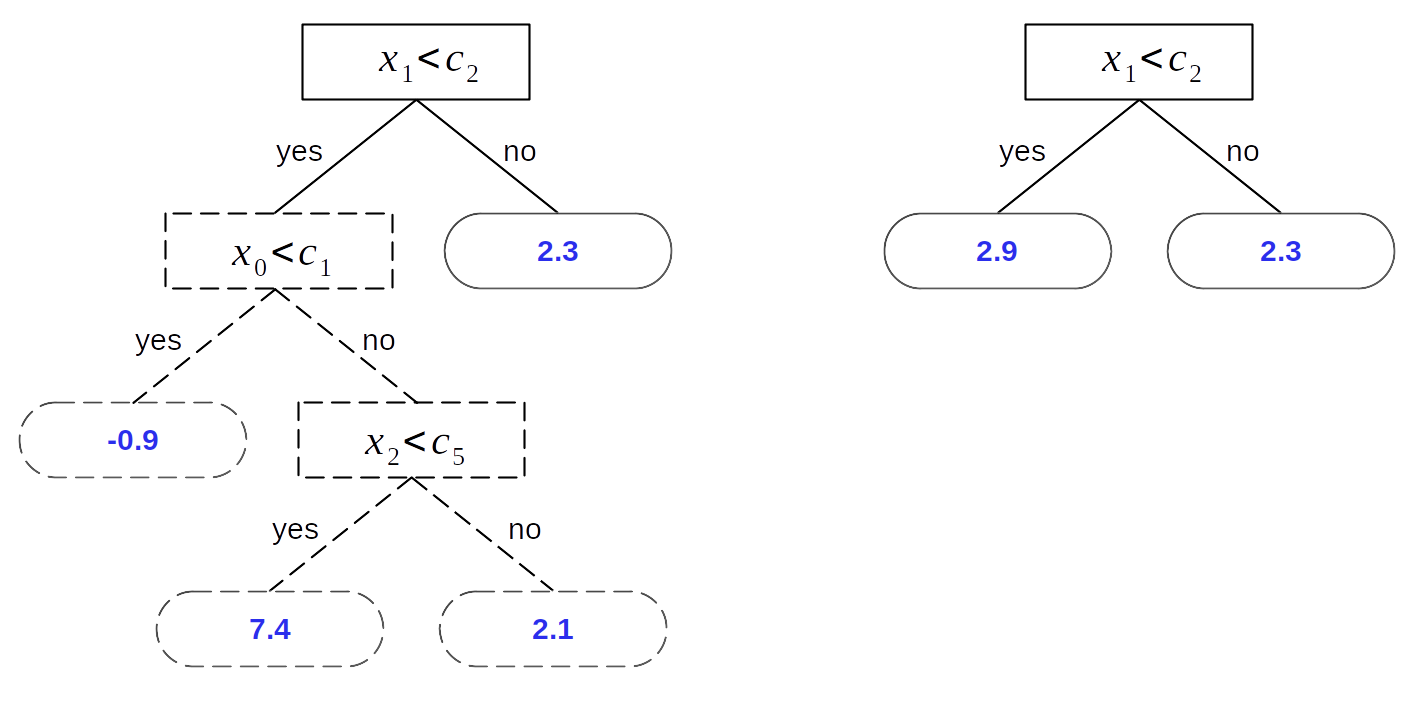}
     \caption{An example of a pruned tree during the computation of a restricted model. The tree on the left represents a tree obtained for the full model. The tree on the right represents the tree that will be used to compute a restricted model that does not include the variable $X_0$.}
     \label{fig:pruned_tree}
\end{figure}

 \texttt{pmb.plot\_variable\_importance} returns two subplots (see Figure~\ref{fig:pearson_bikes}). The top one shows the values of the relative variable importance, estimated as the count of each variable across all sampled trees and then normalized so the sum of all variables' importance is 1. This is a standard plot in the BART literature. The bottom subpanel shows the square of the Pearson correlation coefficient\footnote{also known as the coefficient of determination} between the full model with all covariates and models with fewer covariates. It can be used to find the minimal model capable of making predictions that are close to the full model. To generate this plot, we make two important approximations in order to reduce the computational cost:

\begin{itemize}
    \item We do not evaluate all possible combinations of variables, we simply add one variable at a time, following their relative importance.
    \item We do not refit the model for 1 to n components; instead, we approximate the effect of removing variables by traversing trees from the posterior distribution computed for the full-model and pruning branches without the variable of interest, see Figure~\ref{fig:pruned_tree}. This is similar to the procedure to compute the partial dependence plots, with the difference that for the plots we excluded all but one variable, and for the variable importance we start by excluding all but the most important one, then all but the two most important ones, until we include all variables.
\end{itemize}

\section{Examples}
\label{sec:examples}

In this section, we use a few examples to show how to build PyMC model with BART random variables. We are assuming the reader is already familiar with the PyMC syntax. If that's not the case we recommend readers spend some time reading PyMC's official documentation \url{https://www.pymc.io}.

\subsection{Bikes}

For our first example, we will use a dataset from the University of California Irvine’s Machine Learning Repository \url{https://archive.ics.uci.edu/ml/datasets/bike+sharing+dataset}. As the response variable, we use the number of bikes rented per hour. And for the covariates we use the hour of the day, the temperature, the humidity, and the speed of the wind. 

\begin{figure}[t!]
    \centering
    \includegraphics[width=0.8\textwidth]{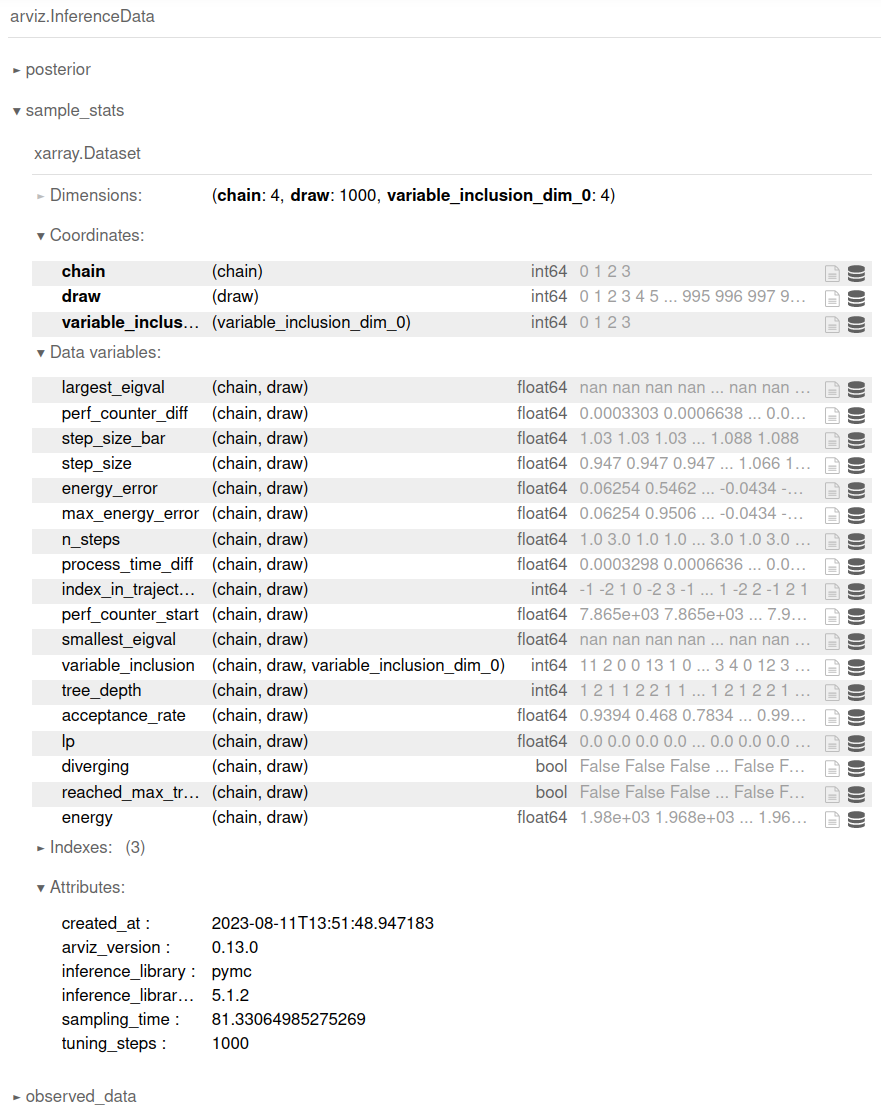}
     \caption{Inference data obtained from  \texttt{model\_bikes}.}
     \label{fig:inf_data}
\end{figure}

The proposed model is:
\begin{listing}[!ht]
\begin{minted}{python}
with pm.Model() as model_bikes:
    α = pm.Exponential("α", 0.1)
    μ = pmb.BART("μ", X, np.log(Y), m=50)
    y = pm.NegativeBinomial("y", mu=np.exp(μ), alpha=α, observed=Y)
    idata_bikes = pm.sample(tune=2000, draws=2000)
\end{minted}
\caption{PyMC model for the bikes example. We have followed the import conventions \texttt{import pymc as pm} and \texttt{import pymc\_experimental as pmx}.}
\label{code:bikes}
\end{listing}
Where we have followed the import conventions  \texttt{import pymc as pm} and  \texttt{import pymc-bart as pmb}. As was mentioned, the posterior samples are stored in ArviZ's InferenceData object~\citep{Kumar2019, arviz}. Figure~\ref{fig:inf_data} shows an HTML representation of this object.

\begin{figure}[t!]
    \centering
    \includegraphics[width=\textwidth]{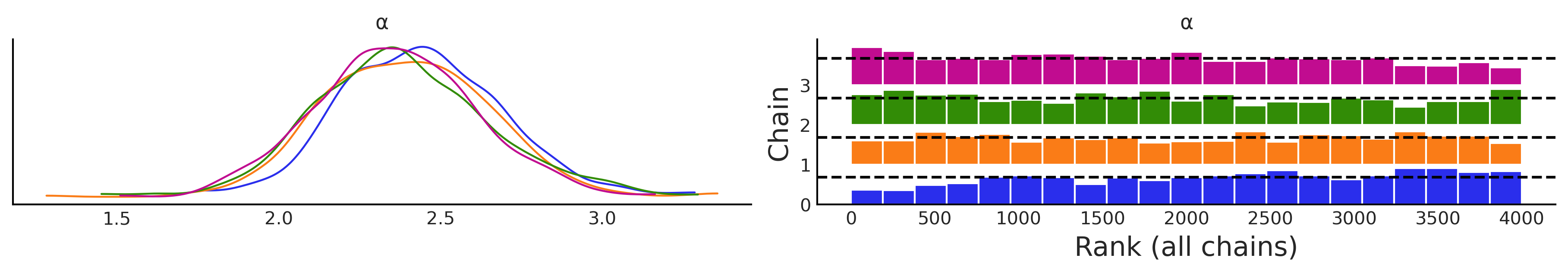}
     \caption{Visual diagnostics for  \texttt{model\_bikes} generated with the command  \texttt{az.plot\_trace(idata\_bikes, var\_names=["α"], kind="rank\_bars")}. On the left,  kernel density estimates computed from the posterior distribution of $\alpha$, and on the right a rank-plot.}
     \label{fig:bikes_diagnostics_non-bart_rv}
\end{figure}

\begin{figure}[t!]
    \centering
    \includegraphics[width=\textwidth]{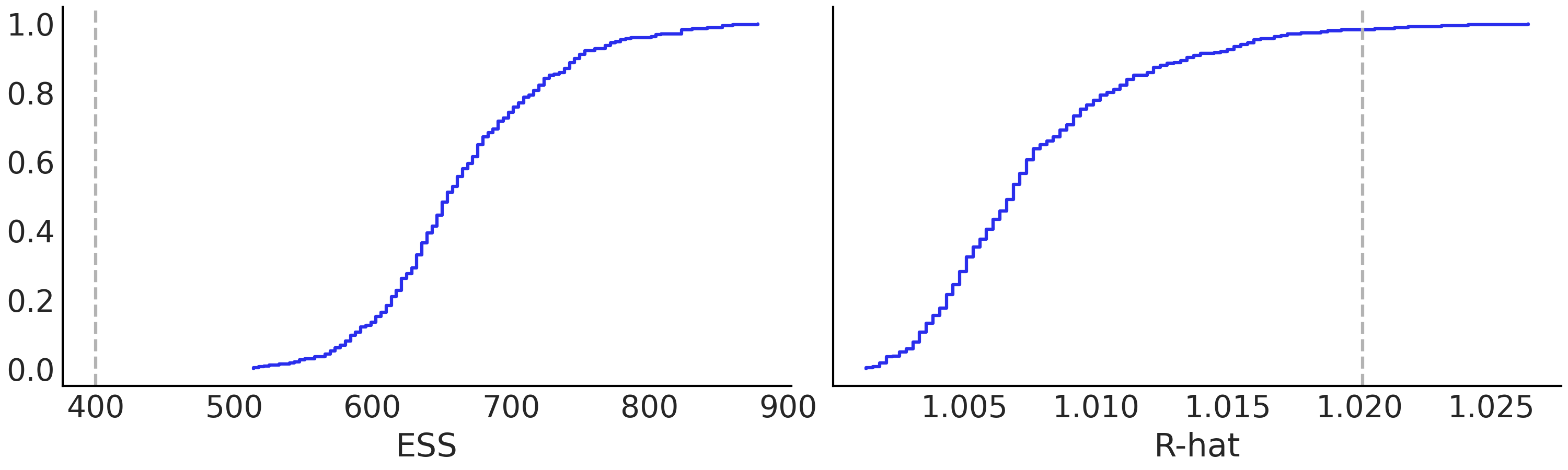}
     \caption{Visual diagnostics for  \texttt{model\_bikes} generated with the command  \texttt{pmb.plot\_convergence(idata\_bikes, var\_names=["μ"]}. On the left is the empirical cumulative distribution function for the effective samples size and on the right for $\hat R$.}
     \label{fig:bikes_diagnostics_bart_rv}
\end{figure}

Figure~\ref{fig:bikes_diagnostics_non-bart_rv} shows a visual diagnostics for $\alpha$ the non-BART random variable in  \texttt{model\_bikes} such plots are a common, and useful, way to visualize sampling convergence. On the other hand, this kind of plot can be less useful to diagnose BART random variables. For that reason PyMC-BART offers the function  \texttt{pmb.plot\_convergence} (see Figure~\ref{fig:bikes_diagnostics_bart_rv}).

From Figure~\ref{fig:bikes_diagnostics_bart_rv} we can see, on the left, that all values of the effective sample size $\mu$ are above the recommended threshold (dashed vertical line) value of 400 (100 per chain). And on the right, we see that most, but not all, of the $\mu$ values, have $\hat R$ below the recommended threshold, taking a few more samples could potentially fix this issue. The value of the threshold for $\hat R$ is computed automatically using a multiple comparison adjustment.

\begin{figure}[t!]
    \centering
    \includegraphics[width=\textwidth]{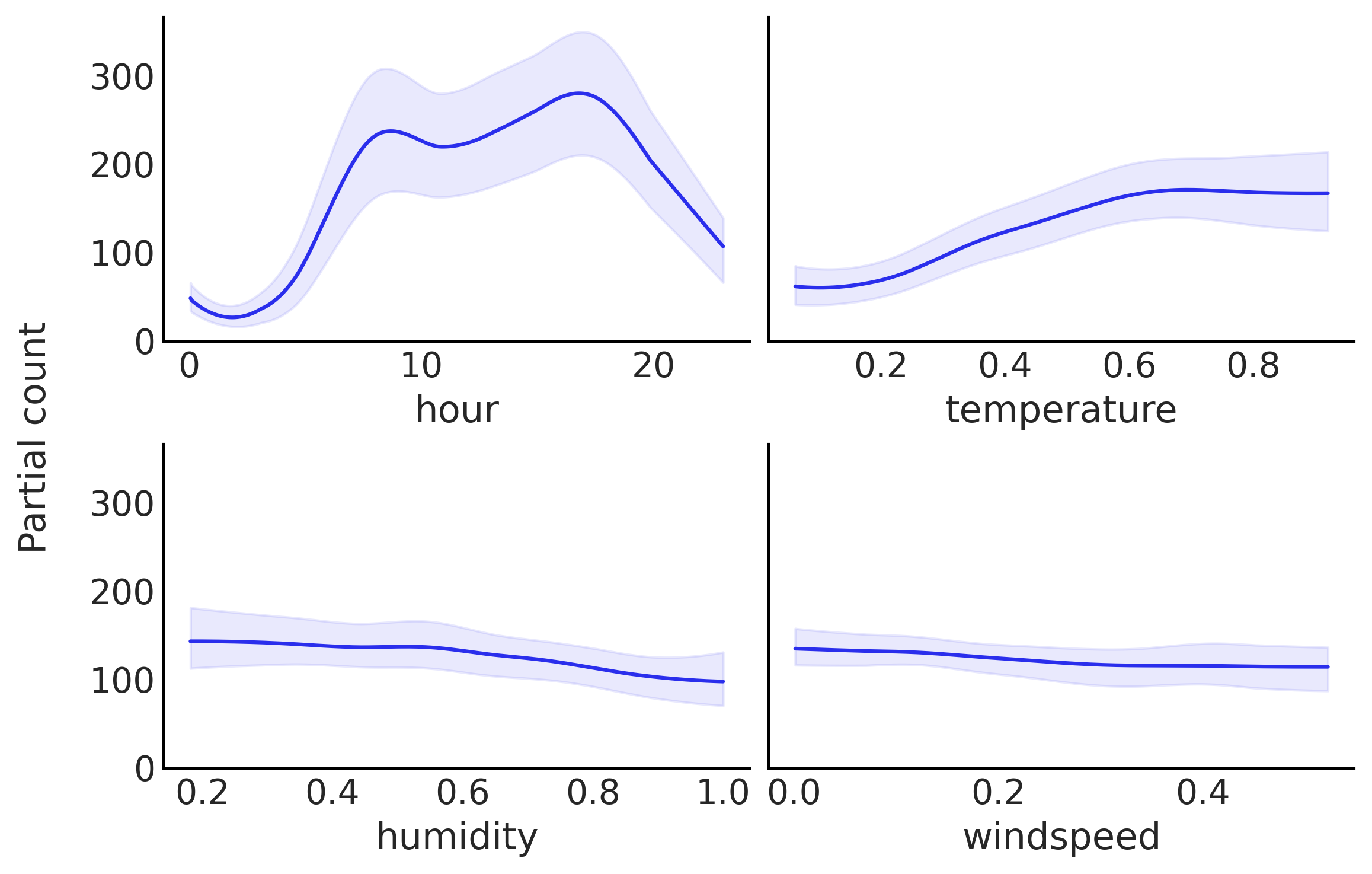}
     \caption{Partial contribution of each variable to the number of rented bikes.}
     \label{fig:pdp_bikes}
\end{figure}

Besides extending PyMC with BART random variables and the PGBART sampler, we also offer a few helper functions, one of which can be used to compute partial dependence plots~\citep{friedman_2001}. For instance, the command   \texttt{pmb.plot\_pdp(idata\_bikes, X=X, Y=Y, grid=(2, 2))} generates Figure~\ref{fig:pdp_bikes}. This allows the user to analyze the partial contribution of each variable. 

From Figure~\ref{fig:pdp_bikes}, we can conclude that:

\begin{enumerate}
    \item The marginal contribution of the variable  \texttt{hour} varies in a more complex way than for the other variables. Starting from a minimum between 0 and 3 hours approximately, it increases to a first peak at around 8; then it decreases slightly to a new, higher peak, at around 17, and finally, it decreases. This pattern can be interpreted as a relationship between working hours and the need to rent bikes.
    \item While the value of  \texttt{temperature} increases, the number of rented bikes also increases, but at some point it stabilizes. This can be explained by saying that at higher temperatures people are more motivated to go out, but there comes a point where this is no longer the case.
    \item  \texttt{humidity} seems to contribute in a very slightly negative way, which could be interpreted as meaning that high humidity does not contribute to people wanting to ride a bike. Nevertheless, the relationship (if any) seems to be tenuous.
    \item   \texttt{windspeed} shows practically no contribution to the motivation to ride a bike.
\end{enumerate}

\begin{figure}[t!]
    \centering
    \includegraphics[width=0.8\textwidth]{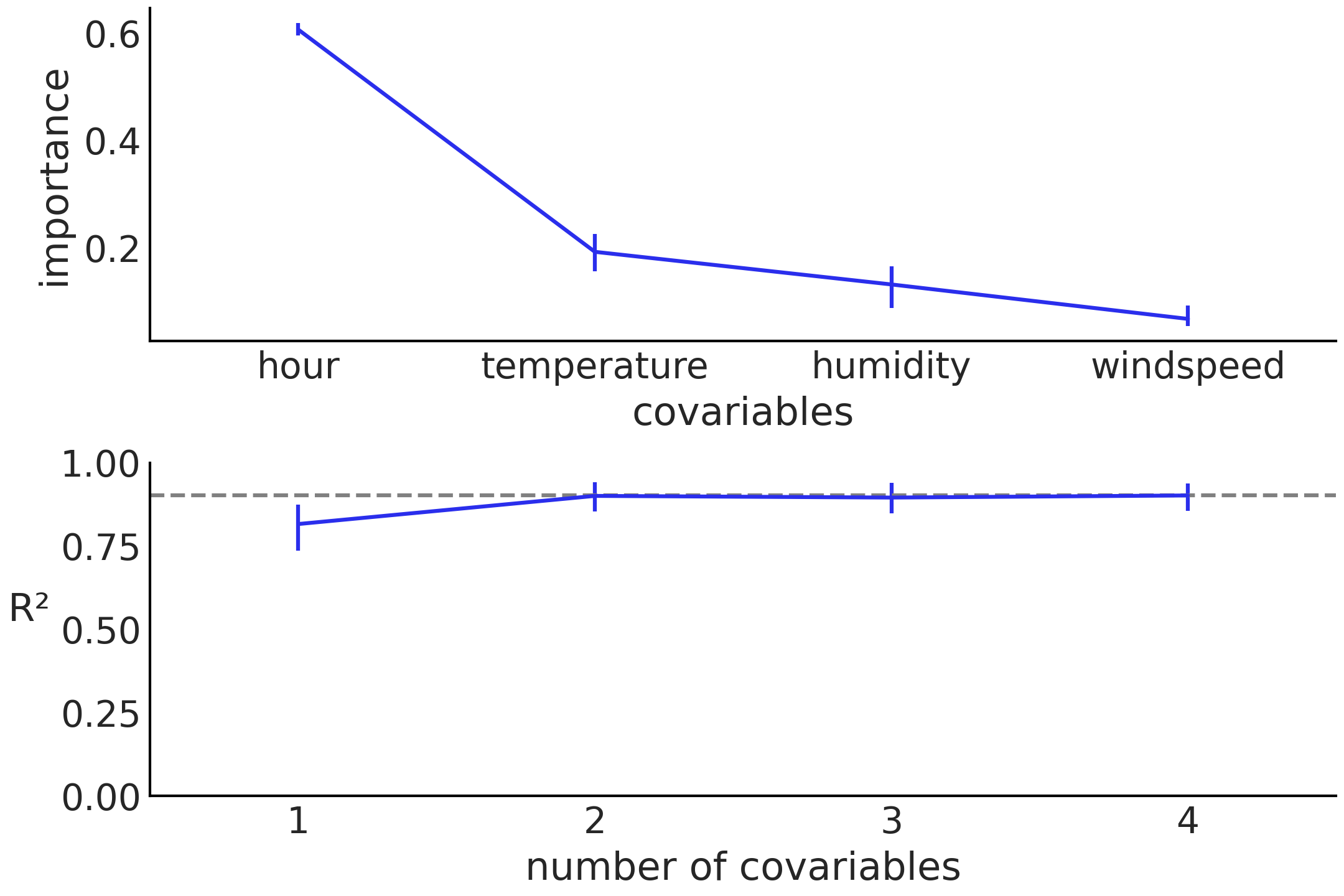}
     \caption{Variable importance results for the Bikes dataset. Top, normalized variable importance for each variable. Bottom, the square of the Pearson correlation coefficient between the predictions made by the full model (all 4 variables are included, gray dash line) and the restricted models (1, 2, 3 or 4 variables included, blue line)}
     \label{fig:pearson_bikes}
\end{figure}

Now we move our focus to the analysis of variable importance. From the top panel of Figure \ref{fig:pearson_bikes} generated using the code \texttt{pmb.plot\_variable\_importance(idata\_bikes, μ, X, samples=100)}

From the top panel, we can see that the variables  \texttt{hour} and  \texttt{temperature} are the most important covariates and that the other two are less important. Notice this is in line with the partial dependence plots from Figure~\ref{fig:pdp_bikes}. This kind of plot is useful to see the relative importance of a variable but is not very useful if we want to select a subset of the variables, as we do not have a clear criterion to separate the variables with “high” importance from those with “low” importance.

In order to provide a variable selection procedure from the computation of  variable importance, we introduce a new plot. We can see an instance in the bottom panel of Figure~\ref{fig:pearson_bikes}. On the x-axis we have the number of variables and on the y-axis the square of the Pearson correlation coefficient between the predictions made by the full-model (all variables included) and the restricted models, i.e., those with only a subset of the variables in the full-model. The variables are included following the relative variable importance order, as shown in the top panel. Thus, in the bikes example, the first variable included is  \texttt{hour}, then  \texttt{hour} and  \texttt{temperature}, followed by  \texttt{hour},  \texttt{temperature} and  \texttt{humidity} and finally the four variables  \texttt{hour},  \texttt{temperature},  \texttt{humidity}, and  \texttt{windspeed}, i.e., the full model. Hence, from Figure~\ref{fig:pearson_bikes} we can see that even a model with a single component,  \texttt{hour} can generate predictions that are very close to the full model. Moreover, the predictions from the model with the two components  \texttt{hour} and \texttt{ temperature} is, on average, indistinguishable from the full model. The error bars represent the 94 \% highest density interval (HDI) of the posterior predictive distribution, that is, of the predictions of the model.

\begin{figure}[t!]
    \centering
    \includegraphics[width=\textwidth]{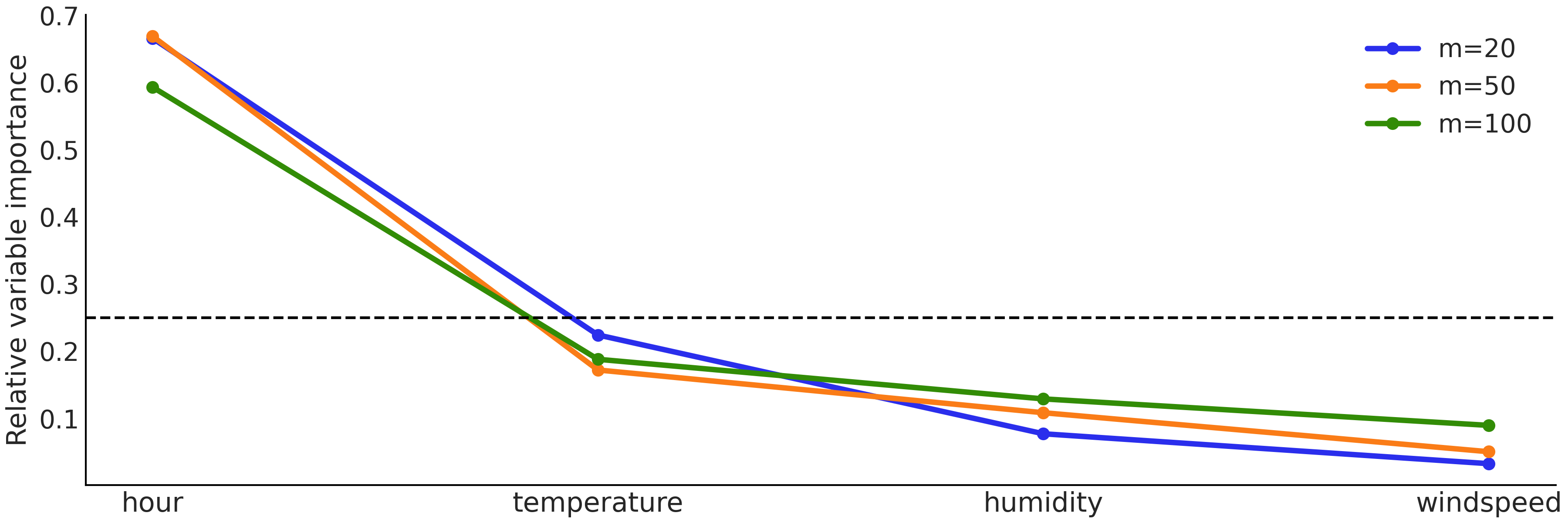}
     \caption{Variable importance for the bike's model. Hour and temperature are the most important covariates. The number of trees  \texttt{m} has little effect on the computed values of the variable importance. The black dashed line represents the uniform importance variable $\left(\frac{1}{4}\right)$.}
    \label{fig:vi_bikes}
\end{figure}

Finally, we close this example by showing that the computation of variable importance is robust with respect to the number of trees \texttt{m}, as can be seen from Figure~\ref{fig:vi_bikes}. We notice this is contrary to the original proposal by~\citet{Chipman2010} where they use a low number of trees (\texttt{m}=20 or 25) for variable importance and a higher one for inference. With PyMC-BART is possible to use the same value of \texttt{m} for both tasks. Hence, our suggestion is to calculate the importance of variables using the same number of trees as employed for inference, typically ranging from 50 to 200 trees. It's important to ensure the adequacy of inference for our specific goals and to verify the absence of convergence issues before proceeding.

\subsection{Friedman function}

The second example uses the Friedman function~\citep{Chipman2010}, which consists of generating data for the random variables $\boldsymbol{X}=(X_0,X_1,...,X_p)$ where $X_0, X_1,...,X_p$ iid $\sim \mathcal{U}(0,1)$ and $Y=f(\boldsymbol{X})+\epsilon=10\sin(\pi X_0 X_1)+20(X_2-0.5)^2+10X_3+5X_4+\epsilon$, where $\epsilon \sim N(0,1)$.

We can see that $Y$ only depends on the first five covariates $\boldsymbol{X}_{0:4}$. Thus, the rest of the covariates $\boldsymbol{X}_{5:p}$ are completely irrelevant. This fact, plus the nonlinearities and interactions, make finding $f(x)$ challenging for standard parametric methods, and thus a good test for BART models.

To fit the data generated from the Friedman function, we use the model:   
\begin{listing}[!ht]
\begin{minted}{python}
    with pm.Model() as model_friedman:
        σ = pm.HalfNormal('σ', 1)
        μ = pmb.BART('μ', X, Y, m=200) 
        y = pm.Normal('y', μ, σ, observed=Y)
        idata_friedman = pm.sample()
\end{minted}
\caption{PyMC model for the Friedman example. This model is essentially the same as that of Code Block \ref{code:bikes}.}
\label{code:friedman}
\end{listing}
We fit this model four times, changing at each instance the number of features (columns) in variable  \texttt{X}, $p\in \{5, 10, 100, 1000\}$. That is, we evaluate BART for an increasing number of irrelevant features.

\begin{figure}[t!]
    \centering
    \includegraphics[width=0.8\textwidth]{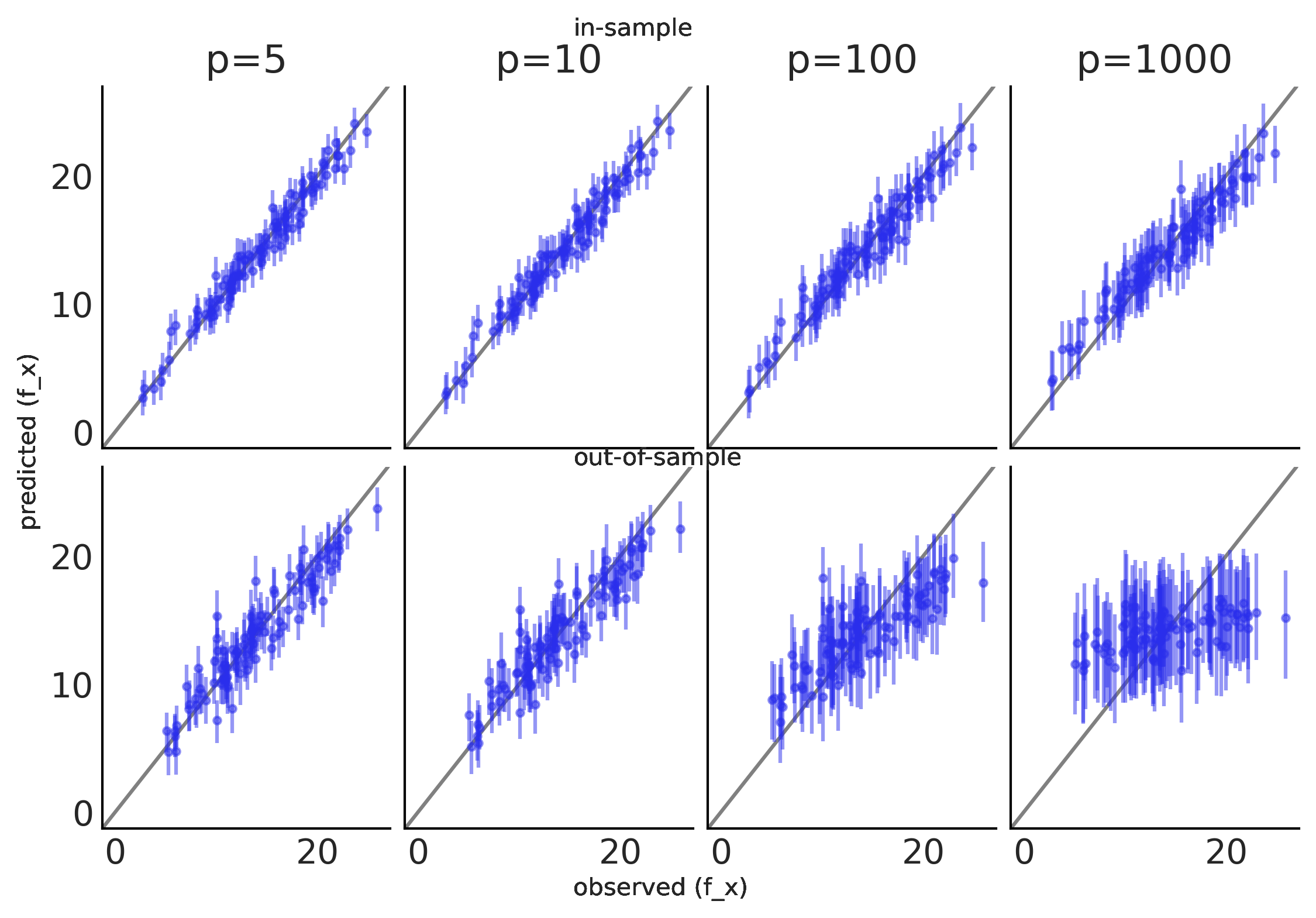}
     \caption{Inference of Friedman  function with dimension $p\in\{5, 10, 100, 1000\}$.}
     \label{fig:friedman_p}
\end{figure}

Figure~\ref{fig:friedman_p} shows the correlation between predicted and observed data with different numbers of covariables $p\in\{5, 10, 100, 1000\}$, the true values are in the x-axis while the y-axis contains the in-sample predictions (top panel) and out-of-sample (bottom panel). The error bars represent the 90\% HDI. The more closely the predictions are to the true function, the closer they will be to the black line at 45$^{\circ}$. For the in-sample predictions, we can see a very good agreement between predicted and observed data even when the number of irrelevant features is much larger than the relevant ones. As expected, out-of-sample predictions are worse than in-sample ones. When the number of irrelevant features is relatively high PyMC-BART predictions are not-robust enough to the number of irrelevant features as previously observed~\cite{linero2018bayesian} with a sparsifying prior. Thus,  our recommendation for when the number of irrelevant covariates is very large compared to the relevant ones, like when p=100 or p=1000 in this example, is to first check variable importance and if the non-relevant variables represent a very large fraction of the total number of covariates, then do a second run keeping only the most relevant covariates. If for some reason the user doesn't want to remove variables from the model then increasing the number of tuning steps and the number of particles can help to improve the out-of-sample fit, at a higher computational cost.

\begin{figure}[t!]
    \centering
    \includegraphics[width=\linewidth]{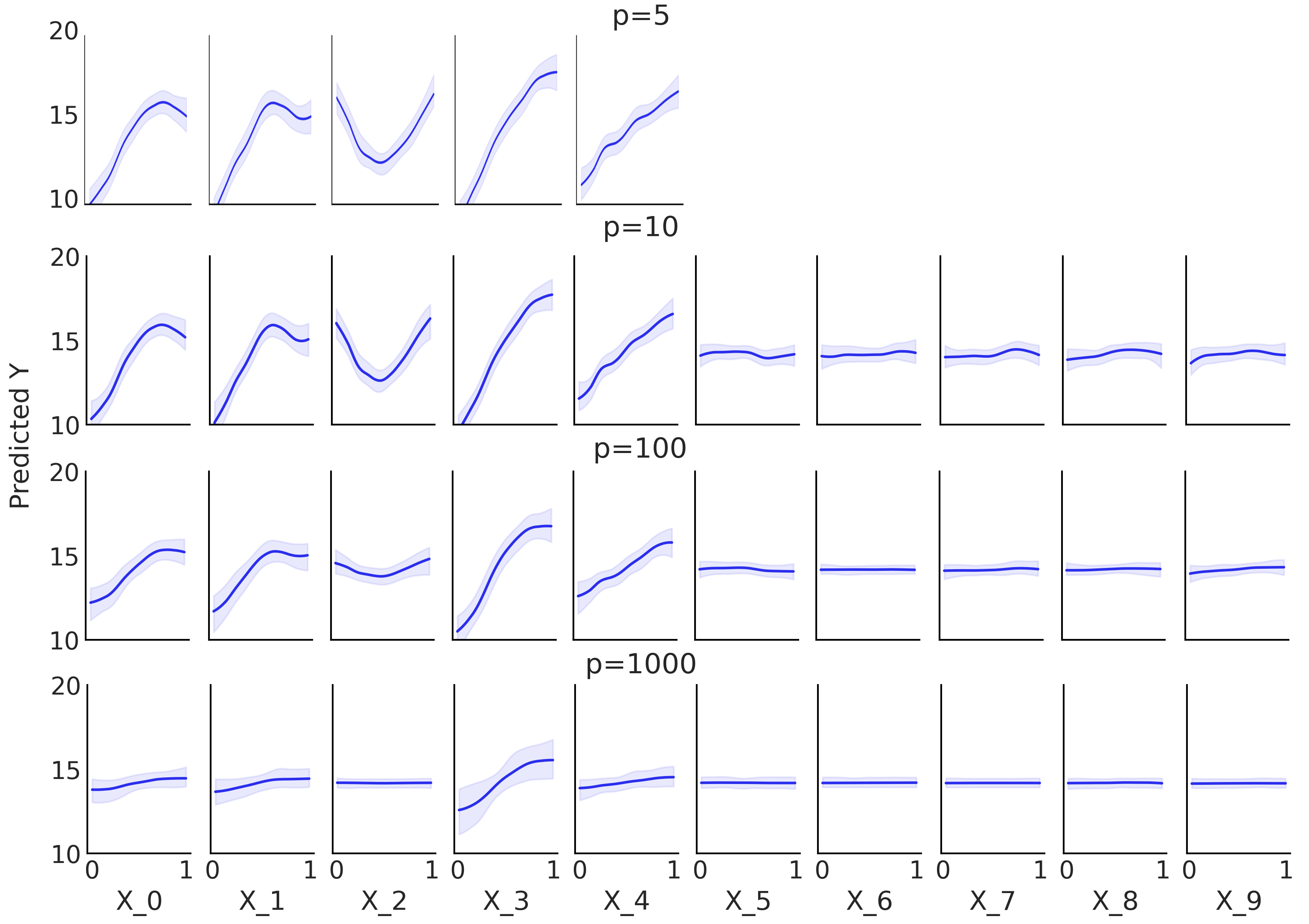}
     \caption{Partial dependence plots with a number of predictors $p\in \{5, 10, 100, 1000\}$.}
     \label{fig:pdp_friedman}
\end{figure}

Figure~\ref{fig:pdp_friedman} shows a partial dependence plot for the BART model  \texttt{model\_friedman}, with different number of covariables $p\in \{5, 10, 100, 1000\}$. We can see that the variables $\boldsymbol{X}_{5:10}$ have almost zero contribution to the response variable $Y$\footnote{With the rest of the variables $\boldsymbol{X}_{10:p}$, not shown here but following the same flat pattern.}, while the first 5 variables $\boldsymbol{X}_{0:4}$ have a larger effect on $Y$. We can also see that, as we increase $p$, the trends are quite robust, although the response becomes flatter. This effect is more clear for $X_2$, and especially for $p=1000$.

\begin{figure}[t!]
    \centering
    \includegraphics[width=0.75\textwidth]{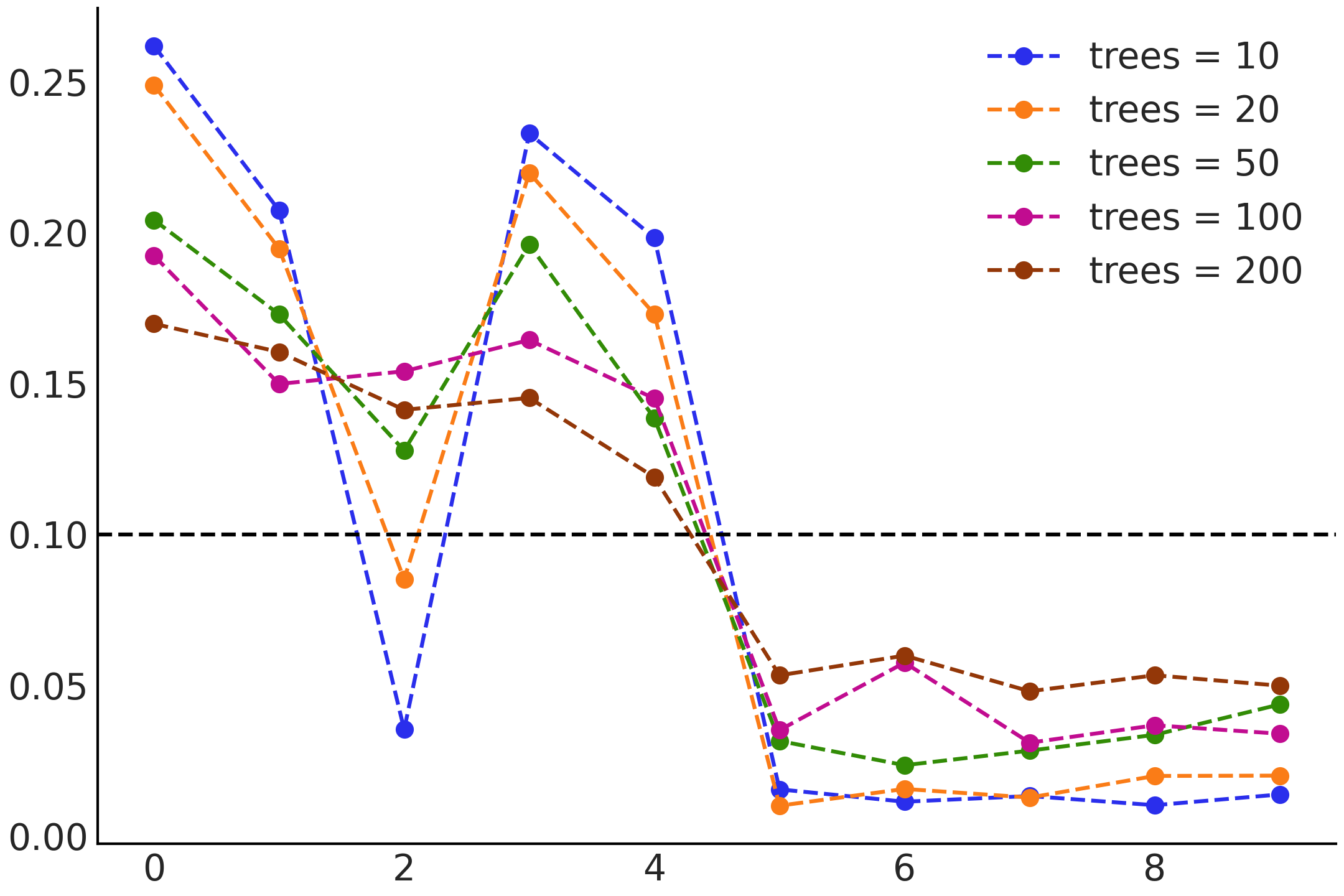}
     \caption{Relatively variable importance of $\boldsymbol{X}_{0:9}$ for number of trees  \texttt{m} $\in\{10, 20, 50, 100, 200\}$. The values are normalized so that the total variable importance sums up to 1. The black dashed line represents the uniform importance variable $\left(\frac{1}{10}\right)$.}
    \label{fig:vi_friedman}
\end{figure}

Next, we compute variable importance, using $p= 10$ and \texttt{model\_friedman}. We observe in Figure~\ref{fig:vi_friedman} that the first five variables $\boldsymbol{X}_{0:4}$ are the more important ones. This is expected from the construction of $Y$, as we already mentioned that the variables $\boldsymbol{X}_{5:p}$ are unrelated to the response variable $Y$. This is in agreement with Figure~\ref{fig:pdp_friedman}.
Additionally, from Figure~\ref{fig:vi_friedman} we can see that variable importance is virtually insensitive to the values of  \texttt{m} when  \texttt{m} >= 50. With fewer trees, 10 or 20, the values of the importance variables have more dispersion, but even in that situation, the  first 5 variables are the most important ones. Again, we can see a qualitative agreement with Figure~\ref{fig:pdp_friedman}.

Our results are similar to that of~\citet {Chipman2010}, with the important difference that, in our case, the variables $\boldsymbol{X}_{5:p}$ have even less importance (almost nil), which is a better result.

\begin{figure}[t!]
    \centering
    \includegraphics[width=0.75\textwidth]{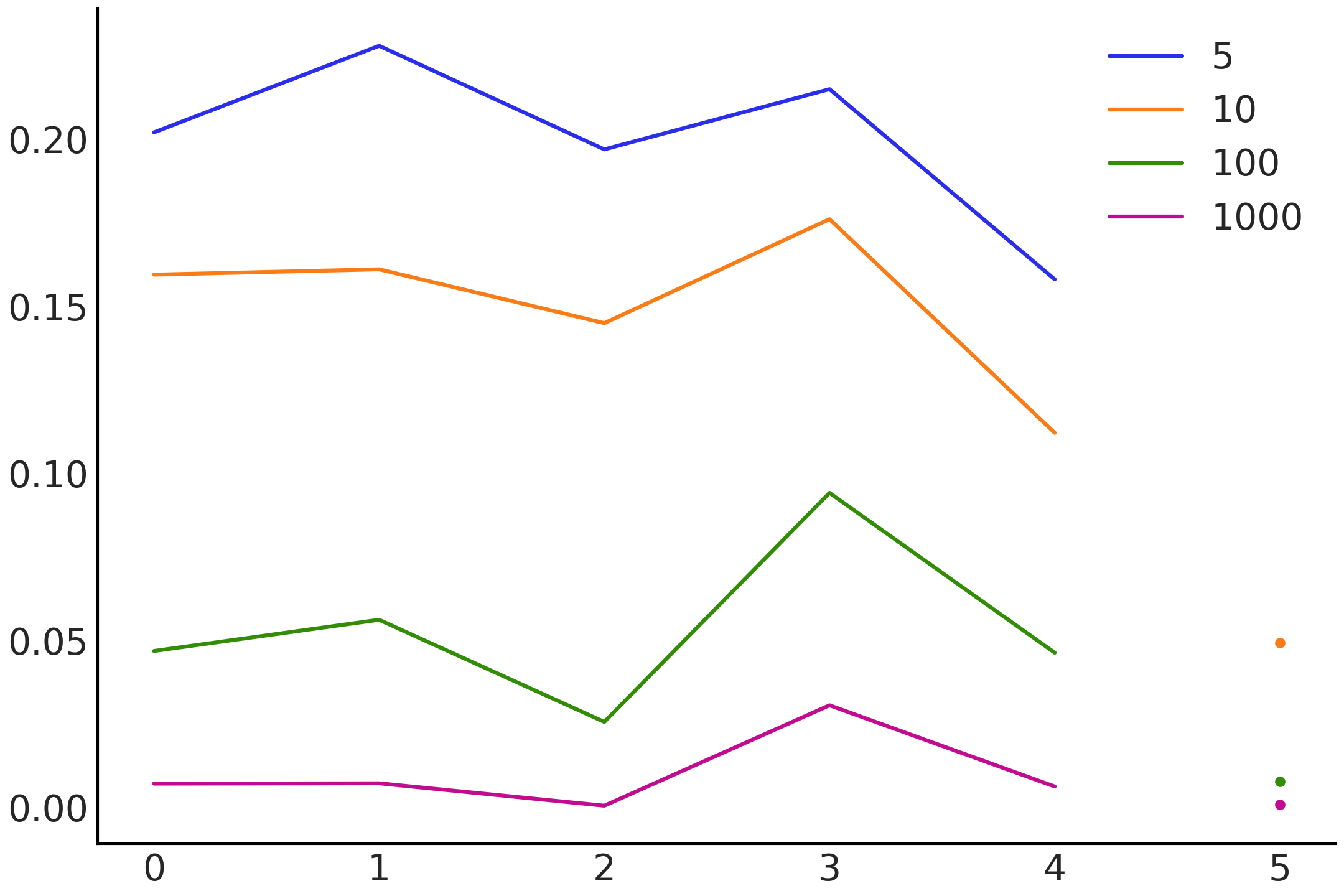}
     \caption{Relative variable importance for $p\in \{5, 10, 100, 1000\}$. The values are normalized so that the total variable importance sums up to 1. The mean of the importance of the variables $\boldsymbol{X}_{5:p}$ is represented with dots.}
    \label{fig:friedman_vi}
\end{figure}

In Figure~\ref{fig:friedman_vi} we can observe that, as the number of irrelevant features increases, the relative importance of the first five covariates decreases, as expected, because the total importance of 1 has to be distributed among more covariates. But we can also see that the relative importance is robust with respect to an increase in the number of non-relevant covariates, because on average the variable importance for the non-relevant covariates is smaller than for the relevant ones.

\begin{figure}[t!]
    \centering
    \includegraphics[width=\textwidth]{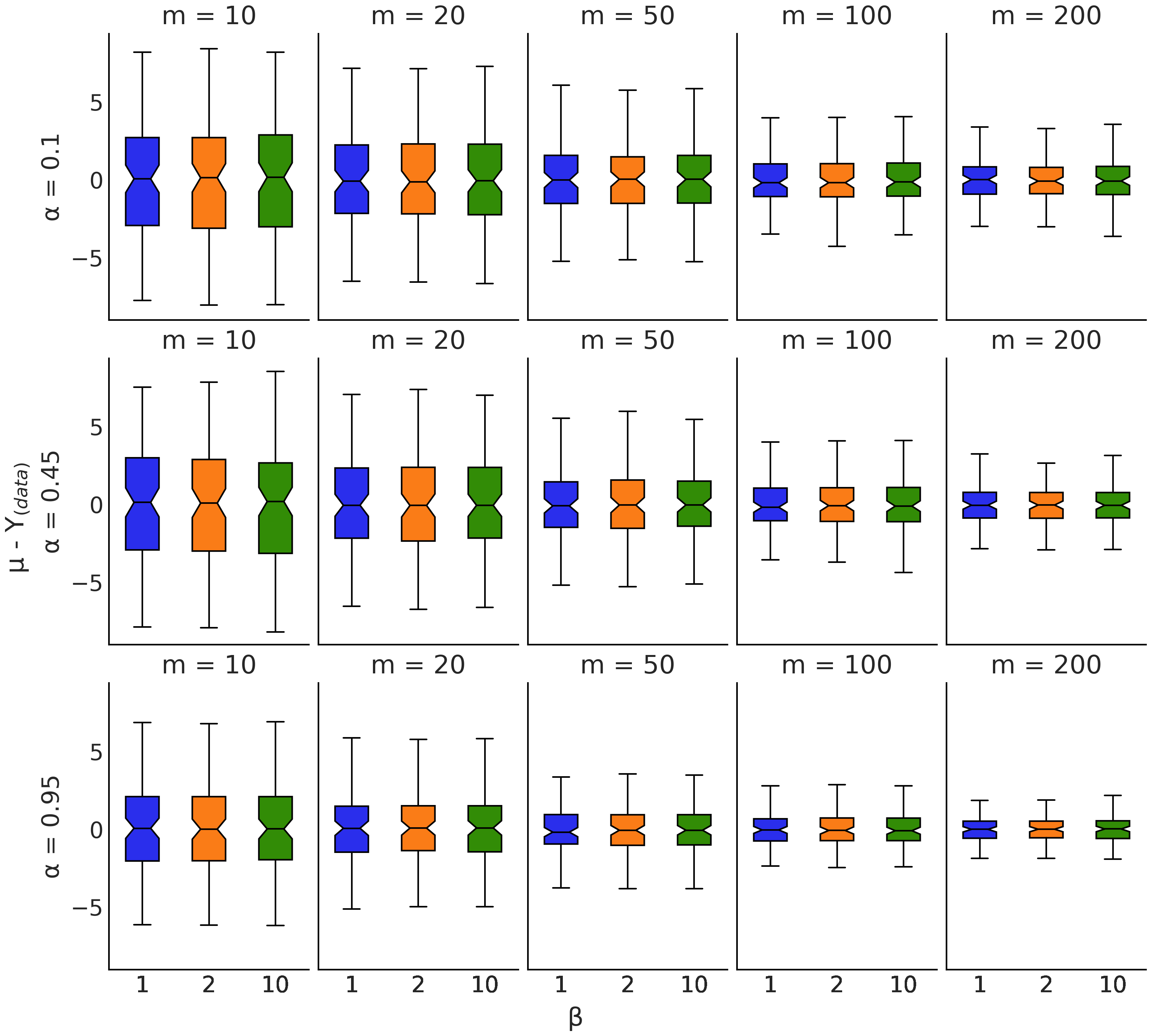}
     \caption{Box plots for the difference between the true Friedman function and the predictions for $m \in \{10, 20, 50, 100, 200\}$ in the panels from left to right.  And $\alpha \in \{0.1, 0.45, 0.95\}$ in the top, middle and bottom row respectively.  In each panel,  $\beta \in \{1, 2, 10\}$ is plotted in blue, orange, and green, respectively}
    \label{fig:boxplot_friedman_i2}
\end{figure}

To evaluate the impact of the hyperparameters  \texttt{alpha} and  \texttt{beta}, which controls the prior for the depth of the trees, we use  \texttt{model\_friedman} for $p=10$, iterating through  \texttt{m},  \texttt{alpha} and  \texttt{beta}, for $\texttt{m} \in \{10, 20, 50, 100, 200\}$,  \texttt{alpha} $\in \{0.1, 0.45, 0.95\}$ and  \texttt{beta} $\in \{1, 2, 10\}$.
In Figure~\ref{fig:boxplot_friedman_i2} we can see that the effect of  \texttt{alpha} parameter and  \texttt{beta} parameter is overall small in comparison with the effect of $m$.
We repeat these calculations for  \texttt{models\_bikes},  \texttt{models\_coal}, and the space influenza toy model with similar outcomes. Because deeper trees are needed to represent higher-order interactions, we created two modified versions of Friedman functions by adding, to the first term, one and two more interactions. Again, the effect of changing  \texttt{alpha} is small.  All results are available at \url{https://github.com/Grupo-de-modelado-probabilista/BART/tree/master/experiments}.

After this experiment and following \citet{Chipman2010}, we decided to set  \texttt{alpha=0.95} and  \texttt{beta=2} as the default values.

\subsection{Cox processes}

We now show how to use BART for a 1D discretized non-homogeneous Poisson process. We use the classical coal mining disaster dataset. The same example, but using Gaussian Processes, can be found in~\citet{martin_bayesian_2018} and the documentation of GPstuff package~\citep{vanhatalo_gpstuff_2013}.

The data consist of timestamps for when the disaster occurred. To be able to fit this data using a regression model, we bin the data as shown next
\begin{listing}[!ht]
\begin{minted}{python}
years = int(coal.max() - coal.min())
bins = years // 4
hist, x_edges = np.histogram(coal, bins=bins)
x_centers = x_edges[:-1] + (x_edges[1] - x_edges[0]) / 2
x_data = x_centers[:, None]
y_data = hist
\end{minted}
\caption{Preprocessing of the coal dataset.}
\label{code:coal_pre}
\end{listing}

where the values of  \texttt{X} correspond to the date of the disasters and  \texttt{Y} are the number of accidents for that date. Because this is a simple function, we use  \texttt{m=20}.

The BART model for this example is:
\begin{listing}[!ht]
\begin{minted}{python}
with pm.Model() as model_coal:
    μ_ = pmb.BART("μ_", X=x_data, Y=np.log(y_data), m=20)
    μ = pm.Deterministic("μ", np.exp(μ_))
    y_pred = pm.Poisson("y_pred", mu=μ, observed=y_data)
    idata_coal = pm.sample()
\end{minted}
     \caption{PyMC model for the coal mining dataset. Compared with the model in Code Block \ref{code:bikes}, the main differences are: the use of a Poisson likelihood and exponential inverse link function np.exp(), and a smaller number of trees m=20.}
\end{listing}

The main differences between this model to the ones seen so far are the use of a Poisson likelihood, the exponential inverse link function  \texttt{np.exp()}, and a smaller number of trees  \texttt{m=20}.

\begin{figure}[t!]
    \centering
    \includegraphics[width=0.8\textwidth]{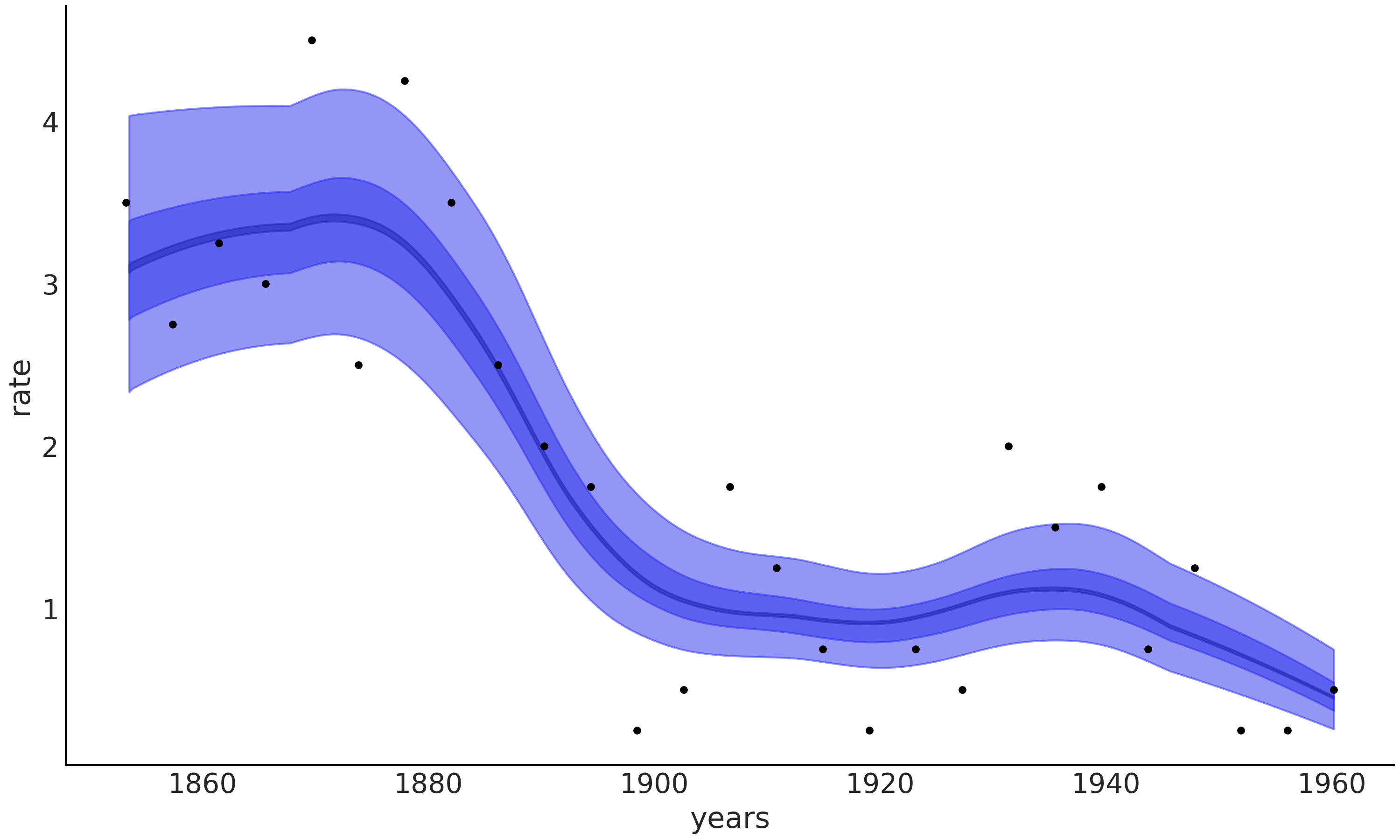}
     \caption{Intensity estimation for the coal mining disaster dataset.}
     \label{fig:coalmining}
\end{figure}

In Figure~\ref{fig:coalmining}, the blue line represents the mean accident rate, and the dark and light blue bands represent the HDI 50\% and 94\% respectively. A notable decrease in accidents can be observed between the years 1880 and 1900.

\subsection{Heteroscedasticity} \label{sec:Heteroscedasticity}

So far we have seen examples of PyMC-BART used to model the mean function, but we can also use it to model a non-constant variance. To exemplify such a scenario, we are going to make use of the marketing dataset \citep{datarium}. We have the budget spent on YouTube advertisement vs the effect on sales. We decided to model the mean as a linear model, with a square root transformation, and let BART be in charge of the standard deviation. The model is:
\begin{listing}[!ht]
\begin{minted}{python}
with pm.Model() as model_marketing:
    α = pm.HalfNormal("α", 50)
    β = pm.HalfNormal("β", 5)
    μ = pm.Deterministic("μ", np.sqrt(α + β * X[:, 0]))
    σ_ = pmb.BART("σ_", X, np.log(Y), m=50)
    σ = pm.Deterministic("σ", np.exp(σ_))
    y = pm.Normal("y", μ, σ, observed=Y)
    idata = pm.sample()
\end{minted}
\caption{PyMC model for the marketing example. Notice, how BART is now used to model the standard deviation
of the Normal response.}
\end{listing}

\begin{figure}
    \centering
    \includegraphics[width=\textwidth]{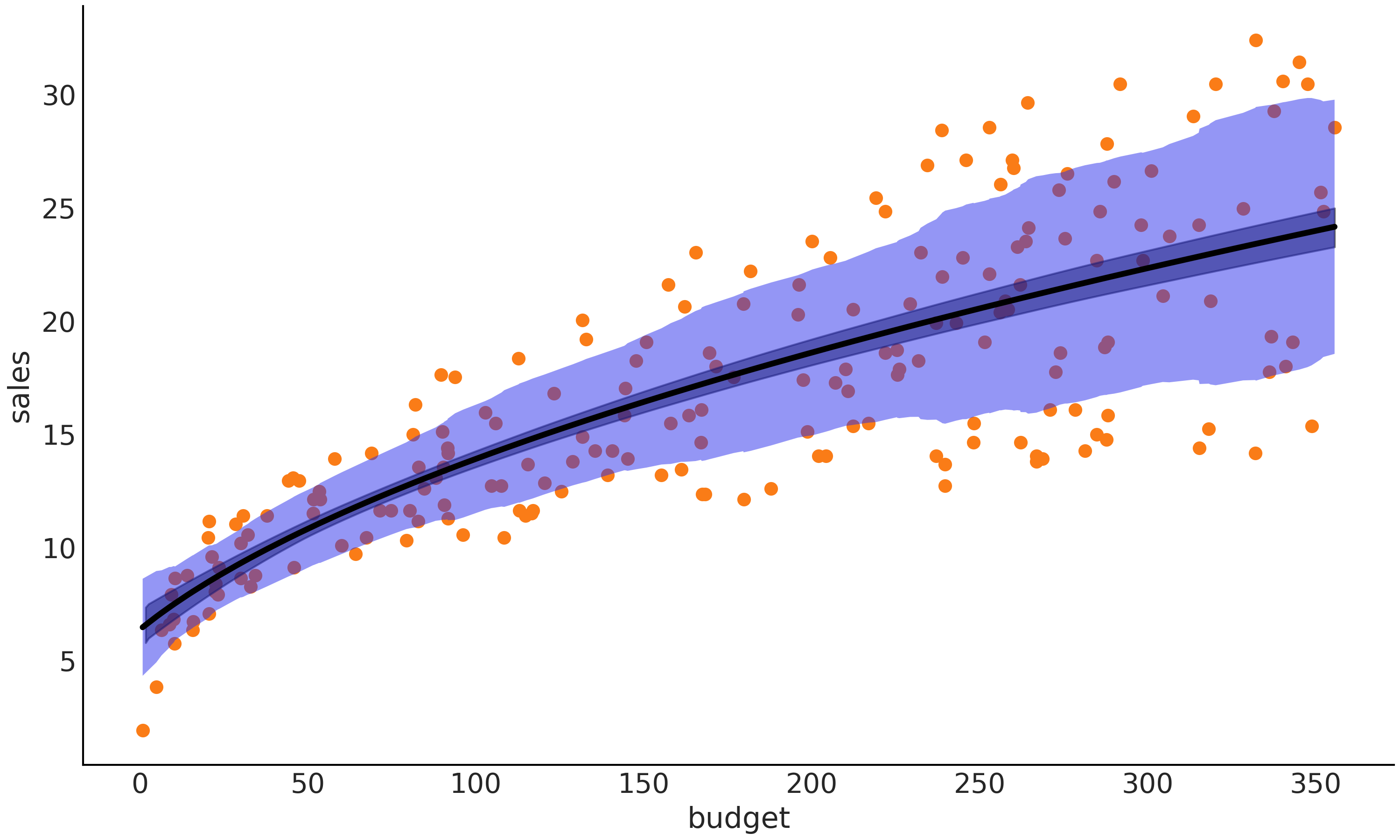}
     \caption{Non-constant variance estimation for the marketing dataset.}
    \label{fig:marketing}
\end{figure}

In Figure~\ref{fig:marketing} we see the mean function as a black line, the 94\% HDI of the mean as a darker blue band, and the 94\% HDI of the standard deviation as a lighter blue band.

Alternatively, we could go fully non-parametric with this example and use BART to model, both the mean and variance. 
\begin{listing}[!ht]
\begin{minted}{python}
with pm.Model() as model_marketing_full:
    w = pmb.BART("w", X, Y, m=200, size=2)
    y = pm.Normal("y", w[0], np.exp(w[1]), observed=Y)
    idata_marketing_full = pm.sample()
\end{minted}
\caption{Fully non-parametric PyMC model for the marketing example. Notice, how BART is defined with a
size = 2 argument and used to model both the mean and standard deviation of the Normal response. Alternative to
size we can pass the shape argument, in this case shape(2, 200), with the second dimension being the sample size.}
\end{listing}
Notice, how BART is defined with a  \texttt{size = 2} argument and used to model both the mean and standard deviation of the Normal response. Alternative to  \texttt{size} we can pass the shape argument, in this case it should be  \texttt{shape(2, 200)}, with the second dimension being the sample size.

\section{Defining the number of trees} \label{sec:loo}

Finding a good value for the number of trees remains important when using BART in practice. Intuitively, as this number controls the flexibility of the BART function, it should be large enough so that the sum of trees is adequate to explain the data, but not so large that the function becomes too flexible. Another practical reason to avoid overshooting  \texttt{m} is the computational cost of BART, both in terms of time and memory, which increases with  \texttt{m}. 

First, \citet{Chipman2010} and others later, have reported that usually the number of trees should be between 20 and 200. For implementations of BART without sparsifying prior over the splitting variables, authors recommend a lower value of  \texttt{m} when computing variable importance than when doing inference. For PyMC-BART we observed that the computation of variable importance is robust with respect to the value of  \texttt{m} (see for example Figures~\ref{fig:vi_bikes} and \ref{fig:vi_friedman}) and in general we recommend using the same value of  \texttt{m} for both inference and variable importance assessment.  

One way to tune  \texttt{m} is to perform K-fold cross-validation, as recommended by~\citet{Chipman2010}. Another option is to approximate cross-validation by using Pareto-smoothed importance sampling leave-one-out cross-validation (PSIS-LOO-CV)~\cite{vehtari_2017, vehtari_2021}. The main advantage of PSIS-LOO-CV is that we only need to fit the model once for each value of  \texttt{m}, instead of K-times as in K-fold cross-validation. It has been reported that PSIS-LOO-CV can lead to overfitting~\cite{Linero2018smoothness} when used to select  \texttt{m} and, hence, we decided to evaluate both PSIS-LOO-CV and 5-fold cross-validation with PyMC-BART. We used these two methods because PSIS-LOO-CV is a fast method to compare models, with strong empirical and theoretical support, and 5-fold cross-validation has been recommended by~\citet{Chipman2010}. Notice that it is not our intention to perform a one-to-one comparison, instead, we are interested in providing practical recommendations for users to pick a value of  \texttt{m}.

For five models and datasets, the 3 simple functions from Figure~\ref{fig:simple_functions}, the models  \texttt{model\_bikes}, the  \texttt{model\_friedman},  \texttt{model\_coal}, and the space influenza model (a toy-model binary classification task taken from \citet{martin_bayesian_2018, Martin2021}) we computed PSIS-LOO-CV as implemented in ArviZ and 5-fold cross-validation. A common pattern we observed when using PSIS-LOO-CV is that, as we increase the value of  \texttt{m} from 10 to 200, in successive steps, the values of the estimated expected log pointwise predictive density (ELPD) keep increasing or, at best,  stop increasing. We observed the same pattern for 5-fold CV, i.e., the root-mean-square deviation (RMSD) decreases with  \texttt{m} or, at best, stabilizes. All the results are available at \url{https://github.com/Grupo-de-modelado-probabilista/BART/tree/master/experiments}.

Regarding these results we want to highlight two facts, the first one is that we observed that even for low values of  \texttt{m} PyMC-BART is able to capture the mean function approximately well and the effect of increasing  \texttt{m} is to refine the fit with a decrease of the residuals. Second, we want to remind the reader that as  \texttt{m} increases, the values of the leaf nodes are shrunk towards zero and, thus, this should  help mitigate overfitting. 

Thus, based on our experiments, and as a rule of thumb, we recommend using values of  \texttt{m} around 50 or less during model building and exploration and switching to values of  \texttt{m} around 200 to obtain the final result. For very simple models and/or small datasets, like the coal mining example in this paper, low values of  \texttt{m} around 50, could be enough. Nevertheless, probably this is not the case for most common scenarios involving smooth functions and datasets around 100 data points or larger. 

\begin{figure}[t!]
    \centering
    \includegraphics[width=0.8\textwidth]{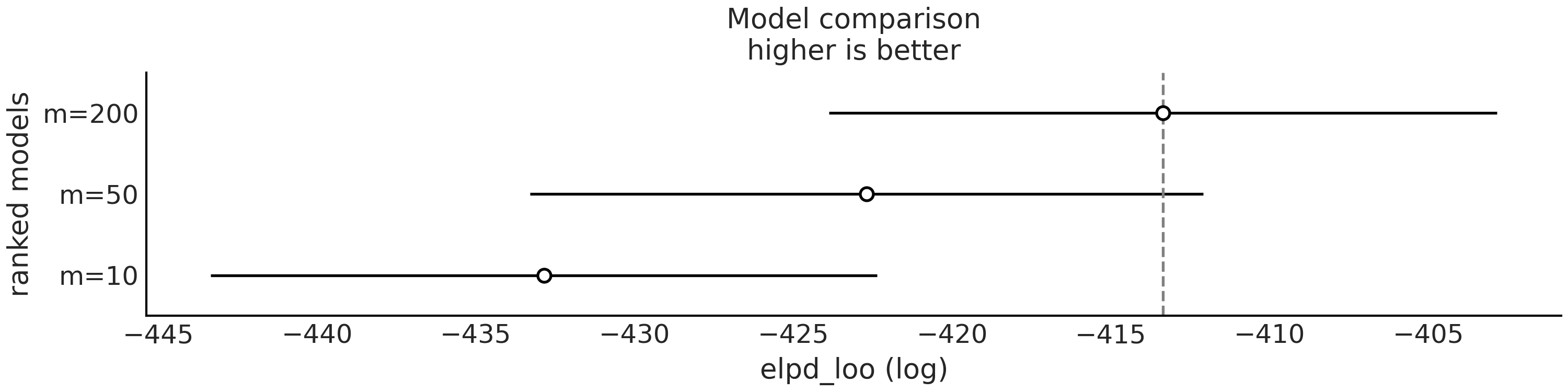}
     \caption{Model comparison using PSIS-LOO-CV for the model and data shown in the first row of Figure~\ref{fig:simple_functions}.}
     \label{fig:LOO_lin}
\end{figure}

If values of  \texttt{m} around 200 become too costly in terms of time and/or memory, so it becomes important to find smaller values for  \texttt{m}, we recommend using PSIS-LOO-CV. Let's say we want to find a suitable value of  \texttt{m} for the model in the first row of Figure~\ref{fig:simple_functions}. We decide to fit the model 3 times with values of m 10, 50, and 100 and hence we will get 3 InferenceData objects, let's call them  \texttt{idata\_sfl\_10, idata\_sfl\_50, idata\_sfl\_200}. To compute LOO and obtain a plot like the one in Figure~\ref{fig:LOO_lin} we can do as follows:
\begin{listing}[!ht]
\begin{minted}{python}
cmp = az.compare({"m=10":idata_sfl_10, 
                  "m=50":idata_sfl_50,
                  "m=200":idata_sfl_100})

az.plot_compare(cmp,
                plot_ic_diff=False, insample_dev=False, legend=False)
\end{minted}
\caption{Example of model comparison using ArviZ to define \texttt{m}.}
\end{listing}

From Figure~\ref{fig:LOO_lin}, we can see that the value of the ELPD follows the order $200 > 50 > 10$, indicating that we should pick  \texttt{m}=200. But it is important to remark that differences equal to or smaller than 4 are considered irrelevant and that, additionally, the uncertainty of the estimated ELPD should also be considered. We can see that there is considerable overlap for  \texttt{m}=50 and  \texttt{m}=200. Thus, picking a value of  \texttt{m}=50, should save time and memory without a practical impact on the fitting.

\section{Discussion and Conclusion} \label{sec:discussion}

PyMC-BART extends PyMC, effectively combining the flexibility of BART models with the power of a modern state-of-the-art probabilistic programming framework. Through a series of empirical evaluations, we have given an overview of how PyMC-BART can be used for inference of non-linear regression models and to assess variable importance or perform variable selection. 

We hope that our contribution will help more users to adopt BART models as part of their toolbox, and we invite researchers and developers to further improve PyMC-BART, as well as to work on similar implementations in other probabilistic programming languages. Some aspects we believe could be improved in the future are: Reducing memory footprint by modifying our tree implementation (or using a Cython or Rust implementation), increasing the efficiency of the sampler (for example we might try to use a better proposal as described in \citet{Pratola_2016}, or rotating splitting rules to explore a wider sample space and improve mixing \citet{Maia_2022}), or add new visualizations like those in \citet{bartviz}. Additionally, we could increase efficiency by modifying the initialization and tuning routine of NUTS conditional to a BART variable in the model. Or explore other distributions for the leaf nodes such as Gaussian Processes \citet{Maia_2022, Wang_2022}, which can reduce the necessary number of trees and improve extrapolation.

\section{Acknowledgments}
This research was supported by the National Agency of Scientific and Technological Promotion ANPCyT, Grant PICT-2018-02212 (O.A.M.) and National Scientific and Technical Research Council CONICET, Grant PIP-0087 to O.A.M. We thank NumFOCUS, a nonprofit 501(c)(3) public charity, for the operational and financial support of PyMC and PyMC-BART. We also want to thank Marcin Elantkowski for reporting a bug in an early version of our code, and also for creating a Rust implementation of PyMC-BART \url{https://github.com/elanmart/rust-pgbart}.

\clearpage
\appendix
\section{Sampling from BART}
\label{sec:pgbart}

We use a sampler inspired by the Particle Gibbs method introduced by~\citet{Lakshminarayanan2015}, but with some modifications in order to be able to define a generalized version of BART, see Algorithm~\ref{alg:pgbart}. We implemented this sampler to work with PyMC, but it is important to note that it can be implemented in other probabilistic programming languages as well. One important consideration is that PGBART is designed to sample trees, and we relied on PyMC's compound step feature to sample non-BART variables. That is, PyMC will automatically assign the PGBART sampler to a  \texttt{pmb.BART} random variable, and if other random variables are present in the model it will assign other samplers to those variables. This makes our implementation of BART flexible enough to easily accommodate a large family of BART models, and even a combination of BART and other models, like, for example, linear regression.

In Algorithm~\ref{alg:pgbart} uppercase letters represent arrays, that is $G$ is an array of $m$ trees, $G_i$ represents the $i$-th tree in the array, and $G_i^{\mu}$ is the output of the $i$-th tree in the array. $ P_j^{l} $ represents the log-likelihood of the $j$-th particle tree.  The first three lines are run once to initialize the algorithm, after that the main loop body (lines 5 to 20) constitutes one step of the sampling algorithm for a BART variable, which can be interleaved with one step of another sampling algorithm, like NUTS, for other variables. To reduce the computational cost at each step, we update a subset of the $m$ trees $m_b \leq m$, by default $m_b$ is 10\% of $m$. To simplify the description of the algorithm, we have defined a few functions, that we now describe. 

\begin{algorithm}
\caption{The PGBART sampler.}\label{alg:pgbart}
\begin{algorithmic}[1]

\Require $X, Y, m, n$ \Comment{n is the number of particles}
\State $G \gets$ \texttt{initialize\_trees}($m$)
\State $\mu$ $\gets \sum_{i=1}^m G_i^{\mu}$ \Comment{compute the sum of trees}
\State $\varepsilon = f(Y)$  \Comment{compute the variance of the leaf nodes from $Y$}
\State $m_b \leq m$
\For{i:=1 to $m_b$}
    \State $\mu_{-i}$ $\gets \mu - G_i^{\mu}$ \Comment{the sum of trees without the tree to replace}
    \State P $\gets$ \texttt{initialize\_particles}($n$)

    \While{trees grow}
        \For{j:=2 to $n$}
            \State $P_j \gets$ \texttt{grow\_tree}($P_j, \mu, \varepsilon, X)$\; \Comment{attempt to grow a particle-tree}
            \State $P_j^{l} \gets \log p(Y \mid \mu_{-i} + P_j^{\mu}, \theta)$ \Comment{compute conditional log-likelihood}
            \EndFor
        \State $\bar W \gets$ \texttt{normalize\_weights}($P_{2:N}$)
        \State $P_{2:N} \gets$\texttt{resample}($P_{2:N}, \bar W$)
    \EndWhile
    \State $\bar W \gets \frac{P_j^{l}}{\sum_j^N P_j^{l}}$ \Comment{use the normalized log-likelihoods as weights}
    \State $G_i \gets$ \texttt{sample}($P, \bar W$) \Comment{sample a new tree from all the particles}
    \State $\mu \gets \mu_{-i}$ + $G_i$
    \State update \texttt{variable\_inclusion}

\EndFor
\end{algorithmic}
\end{algorithm}

\begin{description}
    \item[initialize\_trees]: We set all trees to the mean of $Y$ divided by  \texttt{m}. That is, we set the sum of trees, $\mu$, at the mean of $Y$.
    \item[initialize\_particles]: In order to propose a new $G_i$ tree, we generate $N$ particle-trees. One of these particles is just the tree we want to replace,  $G_i$. This ensures that there is a non-zero probability of keeping the current tree, instead of accepting a new one. The rest of the particles are grown starting from scratch (see  \texttt{grow\_tree} function below) instead of being perturbations of an existing tree. This helps to explore larger regions.
    \item[normalize\_weights]: We normalize the weights, so they are between 0 and 1, and they sum up to 1; this is $\bar W$.  We also compute $w$, which is the sum of unnormalized weights, divided by the number of particles. After resampling, all particles will have the same $w$ weight.
    \item[resample]: Based on, $\bar W$ we use a systematic resampling, of all but the first two particles. This will remove particles with low probability and retain those with higher probability. The number of particles is kept constant, meaning some particles may be repeated.
    \item[sample]: Based on, $\bar W$ we select a single particle and its associated tree.
    \item[variable\_inclusion]: During tuning, this function updates  \texttt{alpha\_vec}, which is the vector with the proportions used to sample the splitting variables. After tuning,   \texttt{alpha\_vec} is fixed, and we instead update the  \texttt{variable\_inclusion} vector, which is then normalized and returned as the estimation of the variable importance. In both cases, the update is done by counting the splitting variables in the returned tree.
    \item[grow\_tree] function attempts to grow a tree based on the following criteria:
    \begin{description}
        \item[node depth]: The probability that a node at depth $d = (0,1,2,...)$ is non-terminal is given by $\alpha(1 + d)^{-\beta}$ with $\alpha \in (0, 1)$ and $\beta \in [0, \infty)$. This prior was proposed and studied by~\citet{Chipman2010}. The default value is  $\alpha = 0.95$ and $\beta =2$.
        \item[splitting variable]: We compute the distribution over the splitting variables from the data. We begin with a flat categorical distribution, $\alpha_{\text{vec}}$, i.e., all covariates have the same chance of being used as splitting variables. During the tuning phase, we continuously update $\alpha_{\text{vec}}$ based on counting the splitting variables in the accepted trees. After the tuning phase, we fix this distribution and use it to sample from the splitting variables.
        \item[splitting values]: Uniform over the observed values.
        \item[leaf values]: We use $\mathcal{N}(\mu_\text{pred}, {\varepsilon^2})$, where $\mu_\text{pred}$ is computed as the mean of the current sum of trees divided by the number of trees  \texttt{m}. $\varepsilon$ is initially computed from $Y$, being $\varepsilon = \frac{3}{\sqrt{m}}$ for binomial data and  $\varepsilon = \frac{Y_\text{std}}{\sqrt{m}}$ for data other than binomial. Because the prior variance on the leaf node parameters can be model dependent during the tuning phase the running variance of the predictions from each fitted tree is computed and used as a proposal. After tuning the variance remains fixed.
    \end{description}

\end{description}

\bibliography{references}

\begin{thebibliography}{46}
\providecommand{\natexlab}[1]{#1}
\providecommand{\url}[1]{\texttt{#1}}
\expandafter\ifx\csname urlstyle\endcsname\relax
  \providecommand{\doi}[1]{doi: #1}\else
  \providecommand{\doi}{doi: \begingroup \urlstyle{rm}\Url}\fi

\bibitem[Abril-Pla et~al.(2023)Abril-Pla, Andreani, Carroll, Dong, Fonnesbeck,
  Kochurov, Kumar, Lao, Luhmann, Martin, Osthege, Vieira, Wiecki, and
  Zinkov]{pymc2023}
O.~Abril-Pla, V.~Andreani, C.~Carroll, L.~Dong, C.~J. Fonnesbeck, M.~Kochurov,
  R.~Kumar, J.~Lao, C.~C. Luhmann, O.~A. Martin, M.~Osthege, R.~Vieira,
  T.~Wiecki, and R.~Zinkov.
\newblock Pymc: A modern and comprehensive probabilistic programming framework
  in python.
\newblock \emph{{PeerJ} Computer Science}, 9:\penalty0 e1516, 2023.
\newblock \doi{10.7717/peerj-cs.1516}.

\bibitem[Bleich et~al.(2014)Bleich, Kapelner, George, and Jensen]{Bleich2014}
J.~Bleich, A.~Kapelner, E.~I. George, and S.~T. Jensen.
\newblock Variable selection for bart: An application to gene regulation.
\newblock \emph{The Annals of Applied Statistics}, 8\penalty0 (3), Sep 2014.
\newblock ISSN 1932-6157.
\newblock \doi{10.1214/14-aoas755}.
\newblock URL \url{http://dx.doi.org/10.1214/14-AOAS755}.

\bibitem[Carlson(2020)]{embarcadero}
C.~J. Carlson.
\newblock embarcadero: Species distribution modelling with bayesian additive
  regression trees in r.
\newblock \emph{Methods in Ecology and Evolution}, 11\penalty0 (7):\penalty0
  850--858, 2020.
\newblock \doi{https://doi.org/10.1111/2041-210X.13389}.
\newblock URL
  \url{https://besjournals.onlinelibrary.wiley.com/doi/abs/10.1111/2041-210X.13389}.

\bibitem[Chen et~al.(2022)Chen, Harhay, Tong, and Li]{chen2022}
X.~Chen, M.~O. Harhay, G.~Tong, and F.~Li.
\newblock A bayesian machine learning approach for estimating heterogeneous
  survivor causal effects: Applications to a critical care trial, 2022.
\newblock URL \url{https://arxiv.org/abs/2204.06657}.

\bibitem[Chipman et~al.(2010)Chipman, George, and McCulloch]{Chipman2010}
H.~A. Chipman, E.~I. George, and R.~E. McCulloch.
\newblock {BART}: {Bayesian} additive regression trees.
\newblock \emph{The Annals of Applied Statistics}, 4\penalty0 (1):\penalty0
  266--298, Mar. 2010.
\newblock ISSN 1932-6157.
\newblock \doi{10.1214/09-AOAS285}.
\newblock URL \url{http://projecteuclid.org/euclid.aoas/1273584455}.

\bibitem[Coltman(2022)]{bartpy}
J.~Coltman.
\newblock Bartpy documentation, 2022.
\newblock URL \url{https://jakecoltman.github.io/bartpy/}.

\bibitem[de~Souza et~al.(2021)de~Souza, Krone-Martins, Carruba,
  de~Cassia~Domingos, Ishida, Alijbaae, Espinoza, and Barletta]{de_Souza_2021}
R.~S. de~Souza, A.~Krone-Martins, V.~Carruba, R.~de~Cassia~Domingos, E.~E.~O.
  Ishida, S.~Alijbaae, M.~H. Espinoza, and W.~Barletta.
\newblock Probabilistic modeling of asteroid diameters from gaia {DR}2 errors.
\newblock \emph{Research Notes of the {AAS}}, 5\penalty0 (8):\penalty0 199, aug
  2021.
\newblock \doi{10.3847/2515-5172/ac205e}.
\newblock URL \url{https://doi.org/10.3847/2515-5172/ac205e}.

\bibitem[Deshpande(2023)]{deshpande2023}
S.~K. Deshpande.
\newblock flexbart: Flexible bayesian regression trees with categorical
  predictors, 2023.

\bibitem[Friedman(2001)]{friedman_2001}
J.~H. Friedman.
\newblock {Greedy function approximation: A gradient boosting machine.}
\newblock \emph{The Annals of Statistics}, 29\penalty0 (5):\penalty0 1189 --
  1232, 2001.
\newblock \doi{10.1214/aos/1013203451}.
\newblock URL \url{https://doi.org/10.1214/aos/1013203451}.

\bibitem[Goldstein et~al.(2013)Goldstein, Kapelner, Bleich, and
  Pitkin]{Goldstein2013PeekingIT}
A.~Goldstein, A.~Kapelner, J.~Bleich, and E.~Pitkin.
\newblock Peeking inside the black box: Visualizing statistical learning with
  plots of individual conditional expectation.
\newblock \emph{Journal of Computational and Graphical Statistics},
  24:\penalty0 44 -- 65, 2013.

\bibitem[He et~al.(2019)He, Yalov, and Hahn]{xbart}
J.~He, S.~Yalov, and P.~R. Hahn.
\newblock Xbart: Accelerated bayesian additive regression trees.
\newblock In K.~Chaudhuri and M.~Sugiyama, editors, \emph{Proceedings of the
  Twenty-Second International Conference on Artificial Intelligence and
  Statistics}, volume~89 of \emph{Proceedings of Machine Learning Research},
  pages 1130--1138. PMLR, 16--18 Apr 2019.
\newblock URL \url{https://proceedings.mlr.press/v89/he19a.html}.

\bibitem[Hill et~al.(2020)Hill, Linero, and Murray]{bart_review}
J.~Hill, A.~Linero, and J.~Murray.
\newblock Bayesian additive regression trees: A review and look forward.
\newblock \emph{Annual Review of Statistics and Its Application}, 7\penalty0
  (1):\penalty0 251--278, 2020.
\newblock \doi{10.1146/annurev-statistics-031219-041110}.
\newblock URL \url{https://doi.org/10.1146/annurev-statistics-031219-041110}.

\bibitem[Hill(2011)]{hill2011bayesian}
J.~L. Hill.
\newblock Bayesian nonparametric modeling for causal inference.
\newblock \emph{Journal of Computational and Graphical Statistics}, 20\penalty0
  (1):\penalty0 217--240, 2011.

\bibitem[Hoffman and Gelman(2014)]{Hoffman2014}
M.~D. Hoffman and A.~Gelman.
\newblock The {No}-{U}-{Turn} {Sampler}: {Adaptively} {Setting} {Path}
  {Lengths} in {Hamiltonian} {Monte} {Carlo}.
\newblock \emph{Journal of Machine Learning Research}, 15\penalty0
  (1):\penalty0 1593--1623, 2014.

\bibitem[Hoyer and Hamman(2017)]{xarray_2017}
S.~Hoyer and J.~Hamman.
\newblock Xarray: {N}-{D} {Labeled} {Arrays} and {Datasets} in {Python}.
\newblock \emph{Journal of Open Research Software}, 5\penalty0 (1), Apr. 2017.
\newblock ISSN 2049-9647.
\newblock \doi{10.5334/jors.148}.

\bibitem[Hu et~al.(2022)Hu, Ji, Ennis, and Hogan]{hu_causal_2022}
L.~Hu, J.~Ji, R.~D. Ennis, and J.~W. Hogan.
\newblock A flexible approach for causal inference with multiple treatments and
  clustered survival outcomes, 2022.
\newblock URL \url{https://arxiv.org/abs/2202.08318}.

\bibitem[Inglis et~al.(2022)Inglis, Parnell, and Hurley]{bartviz}
A.~Inglis, A.~Parnell, and C.~Hurley.
\newblock Visualizations for bayesian additive regression trees, 2022.
\newblock URL \url{https://arxiv.org/abs/2208.08966}.

\bibitem[Kassambara(2019)]{datarium}
A.~Kassambara.
\newblock datarium: {Data} {Bank} for {Statistical} {Analysis} and
  {Visualization}, May 2019.
\newblock URL \url{https://CRAN.R-project.org/package=datarium}.

\bibitem[Kluyver et~al.(2016)Kluyver, Ragan-Kelley, P{\'e}rez, Granger,
  Bussonnier, Frederic, Kelley, Hamrick, Grout, Corlay, Ivanov, Avila, Abdalla,
  and Willing]{Kluyver2016}
T.~Kluyver, B.~Ragan-Kelley, F.~P{\'e}rez, B.~Granger, M.~Bussonnier,
  J.~Frederic, K.~Kelley, J.~Hamrick, J.~Grout, S.~Corlay, P.~Ivanov, D.~Avila,
  S.~Abdalla, and C.~Willing.
\newblock Jupyter notebooks -- a publishing format for reproducible
  computational workflows.
\newblock In F.~Loizides and B.~Schmidt, editors, \emph{Positioning and Power
  in Academic Publishing: Players, Agents and Agendas}, pages 87 -- 90. IOS
  Press, 2016.

\bibitem[Kropat et~al.(2015)Kropat, Bochud, Jaboyedoff, Laedermann, Murith,
  Palacios, and Baechler]{kropat2015improved}
G.~Kropat, F.~Bochud, M.~Jaboyedoff, J.-P. Laedermann, C.~Murith, M.~Palacios,
  and S.~Baechler.
\newblock Improved predictive mapping of indoor radon concentrations using
  ensemble regression trees based on automatic clustering of geological units.
\newblock \emph{Journal of Environmental Radioactivity}, 147:\penalty0 51--62,
  2015.

\bibitem[Kuhn and Johnson(2013)]{kuhn_applied_2013}
M.~Kuhn and K.~Johnson.
\newblock \emph{Applied {Predictive} {Modeling}}.
\newblock Springer-Verlang, New York, 1st ed. 2013, corr. 2nd printing 2018
  edition edition, May 2013.
\newblock ISBN 978-1-4614-6848-6.

\bibitem[Kumar et~al.(2019)Kumar, Carroll, Hartikainen, and Martin]{Kumar2019}
R.~Kumar, C.~Carroll, A.~Hartikainen, and O.~A. Martin.
\newblock Arviz a unified library for exploratory analysis of {Bayesian} models
  in python.
\newblock \emph{Journal of Open Source Software}, 4\penalty0 (33):\penalty0
  1143, 2019.
\newblock ISSN 2475-9066.
\newblock \doi{10.21105/joss.01143}.

\bibitem[Lakshminarayanan et~al.(2015)Lakshminarayanan, Roy, and
  Teh]{Lakshminarayanan2015}
B.~Lakshminarayanan, D.~M. Roy, and Y.~W. Teh.
\newblock Particle gibbs for bayesian additive regression trees, 2015.
\newblock URL \url{https://arxiv.org/abs/1502.04622}.

\bibitem[Lamprinakou et~al.(2020)Lamprinakou, McCoy, Barahona, Gandy, Flaxman,
  and Filippi]{Lamprinakou_2020}
S.~Lamprinakou, E.~McCoy, M.~Barahona, A.~Gandy, S.~Flaxman, and S.~Filippi.
\newblock Bart-based inference for poisson processes, 2020.
\newblock URL \url{https://arxiv.org/abs/2005.07927}.

\bibitem[Leonti et~al.(2010)Leonti, Cabras, Weckerle, Solinas, and
  Casu]{leonti2010causal}
M.~Leonti, S.~Cabras, C.~S. Weckerle, M.~N. Solinas, and L.~Casu.
\newblock The causal dependence of present plant knowledge on
  herbals—contemporary medicinal plant use in campania (italy) compared to
  matthioli (1568).
\newblock \emph{Journal of Ethnopharmacology}, 130\penalty0 (2):\penalty0
  379--391, 2010.

\bibitem[Li et~al.(2022)Li, Liu, Conaty, Zhu, Moncuquet, Stiller, and
  Wilson]{li_genomic_2022}
Z.~Li, S.~Liu, W.~Conaty, Q.-H. Zhu, P.~Moncuquet, W.~Stiller, and I.~Wilson.
\newblock Genomic prediction of cotton fibre quality and yield traits using
  {Bayesian} regression methods.
\newblock \emph{Heredity}, pages 1--10, May 2022.
\newblock ISSN 1365-2540.
\newblock \doi{10.1038/s41437-022-00537-x}.
\newblock URL \url{https://www.nature.com/articles/s41437-022-00537-x}.
\newblock Publisher: Nature Publishing Group.

\bibitem[Linero(2018)]{linero2018bayesian}
A.~R. Linero.
\newblock Bayesian regression trees for high-dimensional prediction and
  variable selection.
\newblock \emph{Journal of the American Statistical Association}, 113\penalty0
  (522):\penalty0 626--636, 2018.

\bibitem[Linero(2022)]{linero2022}
A.~R. Linero.
\newblock Generalized bayesian additive regression trees models: Beyond
  conditional conjugacy, 2022.
\newblock URL \url{https://arxiv.org/abs/2202.09924}.

\bibitem[Linero and Yang(2018)]{Linero2018smoothness}
A.~R. Linero and Y.~Yang.
\newblock Bayesian regression tree ensembles that adapt to smoothness and
  sparsity.
\newblock \emph{Journal of the Royal Statistical Society B}, 80\penalty0
  (5):\penalty0 1087--1110, 2018.
\newblock \doi{https://doi.org/10.1111/rssb.12293}.
\newblock URL
  \url{https://rss.onlinelibrary.wiley.com/doi/abs/10.1111/rssb.12293}.

\bibitem[Maia et~al.(2022)Maia, Murphy, and Parnell]{Maia_2022}
M.~Maia, K.~Murphy, and A.~C. Parnell.
\newblock Gp-bart: a novel bayesian additive regression trees approach using
  gaussian processes, 2022.
\newblock URL \url{https://arxiv.org/abs/2204.02112}.

\bibitem[Martin(2018)]{martin_bayesian_2018}
O.~Martin.
\newblock \emph{Bayesian {Analysis} with {Python}: {Introduction} to
  {Statistical} {Modeling} and {Probabilistic} {Programming} {Using} {PyMC3}
  and {ArviZ}, 2nd {Edition}}.
\newblock Packt Publishing, 2 edition edition, Dec. 2018.

\bibitem[Martin et~al.(2022)Martin, Hartikainen, Carroll, and Abril-Pla]{arviz}
O.~Martin, A.~Hartikainen, C.~Carroll, and O.~Abril-Pla.
\newblock Arviz, May 2022.
\newblock URL \url{https://doi.org/10.5281/zenodo.6547007}.

\bibitem[Martin et~al.(2021)Martin, Kumar, and Lao]{Martin2021}
O.~A. Martin, R.~Kumar, and J.~Lao.
\newblock \emph{Bayesian Modeling and Computation in Python}.
\newblock {Chapman and Hall/CRC}, Boca Raton, 1st edition, 2021.
\newblock ISBN 978-0-3678-9436-8.

\bibitem[Orlandi et~al.(2021)Orlandi, Murray, Linero, and
  Volfovsky]{Orlandi_2021}
V.~Orlandi, J.~Murray, A.~Linero, and A.~Volfovsky.
\newblock Density regression with bayesian additive regression trees, 2021.
\newblock URL \url{https://arxiv.org/abs/2112.12259}.

\bibitem[Pratola(2016)]{Pratola_2016}
M.~T. Pratola.
\newblock {Efficient Metropolis–Hastings Proposal Mechanisms for Bayesian
  Regression Tree Models}.
\newblock \emph{Bayesian Analysis}, 11\penalty0 (3):\penalty0 885 -- 911, 2016.
\newblock \doi{10.1214/16-BA999}.
\newblock URL \url{https://doi.org/10.1214/16-BA999}.

\bibitem[Pratola et~al.(2020)Pratola, Chipman, George, and
  McCulloch]{pratola_2020}
M.~T. Pratola, H.~A. Chipman, E.~I. George, and R.~E. McCulloch.
\newblock Heteroscedastic {BART} via {Multiplicative} {Regression} {Trees}.
\newblock \emph{Journal of Computational and Graphical Statistics}, 29\penalty0
  (2):\penalty0 405--417, Apr. 2020.
\newblock ISSN 1061-8600.
\newblock \doi{10.1080/10618600.2019.1677243}.
\newblock URL \url{https://doi.org/10.1080/10618600.2019.1677243}.
\newblock Publisher: Taylor \& Francis \_eprint:
  https://doi.org/10.1080/10618600.2019.1677243.

\bibitem[Rockova and Saha(2018)]{Rockova2018}
V.~Rockova and E.~Saha.
\newblock On theory for bart, 2018.
\newblock URL \url{https://arxiv.org/abs/1810.00787}.

\bibitem[Sparapani et~al.(2021)Sparapani, Spanbauer, and
  McCulloch]{bart_r_package}
R.~Sparapani, C.~Spanbauer, and R.~McCulloch.
\newblock Nonparametric machine learning and efficient computation with
  bayesian additive regression trees: The bart r package.
\newblock \emph{Journal of Statistical Software}, 97\penalty0 (1):\penalty0
  1–66, 2021.
\newblock \doi{10.18637/jss.v097.i01}.
\newblock URL
  \url{https://www.jstatsoft.org/index.php/jss/article/view/v097i01}.

\bibitem[Steele and Schwartz(2022)]{Steele2022}
K.~M. Steele and M.~H. Schwartz.
\newblock Causal effects of motor control on gait kinematics after orthopedic
  surgery in cerebral palsy: a machine-learning approach.
\newblock \emph{medRxiv}, 2022.
\newblock \doi{10.1101/2022.01.04.21268561}.
\newblock URL
  \url{https://www.medrxiv.org/content/early/2022/01/05/2022.01.04.21268561}.

\bibitem[Tan and Roy(2019)]{general_bart}
Y.~V. Tan and J.~Roy.
\newblock Bayesian additive regression trees and the general bart model.
\newblock \emph{Statistics in Medicine}, 38\penalty0 (25):\penalty0 5048--5069,
  2019.
\newblock \doi{https://doi.org/10.1002/sim.8347}.
\newblock URL \url{https://onlinelibrary.wiley.com/doi/abs/10.1002/sim.8347}.

\bibitem[Vanhatalo et~al.(2013)Vanhatalo, Riihimäki, Hartikainen, Jylänki,
  Tolvanen, and Vehtari]{vanhatalo_gpstuff_2013}
J.~Vanhatalo, J.~Riihimäki, J.~Hartikainen, P.~Jylänki, V.~Tolvanen, and
  A.~Vehtari.
\newblock {GPstuff}: {Bayesian} modeling with {Gaussian} processes.
\newblock \emph{J. Mach. Learn. Res.}, 2013.

\bibitem[Vehtari et~al.(2017)Vehtari, Gelman, and Gabry]{vehtari_2017}
A.~Vehtari, A.~Gelman, and J.~Gabry.
\newblock Practical {Bayesian} model evaluation using leave-one-out
  cross-validation and {WAIC}.
\newblock \emph{Statistics and Computing}, 27\penalty0 (5):\penalty0
  1413--1432, Sept. 2017.
\newblock ISSN 1573-1375.
\newblock \doi{10.1007/s11222-016-9696-4}.
\newblock URL \url{https://doi.org/10.1007/s11222-016-9696-4}.

\bibitem[Vehtari et~al.(2021{\natexlab{a}})Vehtari, Gelman, Simpson, Carpenter,
  and Bürkner]{10.1214/20-BA1221}
A.~Vehtari, A.~Gelman, D.~Simpson, B.~Carpenter, and P.-C. Bürkner.
\newblock {Rank-Normalization, Folding, and Localization: An Improved
  $\widehat{R}$ for Assessing Convergence of MCMC (with Discussion)}.
\newblock \emph{Bayesian Analysis}, 16\penalty0 (2):\penalty0 667 -- 718,
  2021{\natexlab{a}}.
\newblock \doi{10.1214/20-BA1221}.
\newblock URL \url{https://doi.org/10.1214/20-BA1221}.

\bibitem[Vehtari et~al.(2021{\natexlab{b}})Vehtari, Simpson, Gelman, Yao, and
  Gabry]{vehtari_2021}
A.~Vehtari, D.~Simpson, A.~Gelman, Y.~Yao, and J.~Gabry.
\newblock Pareto {Smoothed} {Importance} {Sampling}.
\newblock \emph{arXiv:1507.02646 [stat]}, Feb. 2021{\natexlab{b}}.
\newblock URL \url{http://arxiv.org/abs/1507.02646}.
\newblock arXiv: 1507.02646 version: 7.

\bibitem[Wang et~al.(2022)Wang, He, and Hahn]{Wang_2022}
M.~Wang, J.~He, and P.~R. Hahn.
\newblock Local gaussian process extrapolation for bart models with
  applications to causal inference, 2022.
\newblock URL \url{https://arxiv.org/abs/2204.10963}.

\bibitem[Zhou(2012)]{zhou_ensemble_2012}
Z.-H. Zhou.
\newblock \emph{Ensemble {Methods}: {Foundations} and {Algorithms}}.
\newblock CRC press, Boca Raton, FL, 1 edition edition, June 2012.
\newblock ISBN 978-1-4398-3003-1.

\end{thebibliography}

\end{document}